\documentclass[prb,amsfonts,amssymb,floats,twocolumn,superscriptaddress,aps]{revtex4-2}
\usepackage{amsmath}
\usepackage{float}
\usepackage[caption = false]{subfig}
\usepackage[dvipsnames]{xcolor}
\usepackage{braket}
\usepackage{bm}
\usepackage{dsfont}
\usepackage[export]{adjustbox}
\usepackage{graphicx}
\usepackage{tikz}
\usepackage{placeins}
\usepackage{bbm}
\usepackage{lipsum}
\usepackage[version=4]{mhchem}
\allowdisplaybreaks
\graphicspath{{../Figures/}{./Figures/}}

\newcommand{\br}{\boldsymbol{\rho}}

\begin{document}

\title{Two-triplon excitations of the Kitaev-Heisenberg-Bilayer}
\author{Erik Wagner}
\affiliation{Institute for Theoretical Physics,
Technical University Braunschweig, D-38106 Braunschweig, Germany}
\author{Wolfram Brenig}
\affiliation{Institute for Theoretical Physics,
Technical University Braunschweig, D-38106 Braunschweig, Germany}

\date{\today}

\begin{abstract}
We study the spectrum of a bilayer of Kitaev magnets on the ho\-ney\-comb
lattice, coupled by Heisenberg exchange in the quantum-dimer phase at strong
interlayer coupling. Using the perturbative Continuous Unitary Transformation
(pCUT) we perform series expansion starting from the fully dimerized limit,
to evaluate the elementary excitations, reaching up to and focusing on the
two-triplon sector. In stark contrast to conventional bilayer quantum
magnets, and because of the broken $SU(2)$-invariance, encoded in the
intralayer directional compass-exchange, the bilayer Kitaev magnet is shown
to exhibit a rich structure of two-triplon scattering-state continua, as well
as several collective two-triplon (anti)bound states. Direct physical
pictures for the occurrence of the latter are provided and the (anti)bound
states are studied versus the stacking type, the spin components, and the
exchange parameters. In addition to the two-triplon spectra, we investigate
a corresponding experimental probe by evaluating the magnetic
Raman-scattering intensity. We find a very strong sensitivity of this
intensity on the two-triplon interactions and the scattering geometry,
however a signal from the (anti)bound states appears only in very close
proximity to the continuum.
\end{abstract}

\maketitle

\section{Introduction}
\label{sec:intro}

Quantum phase transitions (QPTs) in local moment systems are of great interest
\cite{Sachdev2008}. In this context models are under intense scrutiny,
comprising lattices of antiferromagnetic dimers, coupled by interdimer exchange
networks. Tuning the interdimer exchange, these models allow for QPTs between
states of weakly interacting singlets and various quantum phases determined by
the type of interdimer network. Paradigmatic examples along this line include
the square-lattice Heisenberg bilayer \cite{Wang2006}, three dimensional dimer
networks \cite{Matsumoto2004}, frustrated triangular and $J_1$-$J_2$ Heisenberg
bilayers \cite{Singh1998,Bishop2019}, as well as the famous orthogonal dimer
model \cite{Miyahara1999}. Each of these models has been suggested to have
corresponding realizations on the materials side, including BaCuSi$_2$O$_6$
\cite{Sebastian2006}, TlCuCl$_3$ \cite{Merchant2014}, Ba$_3$Mn$_2$O$_8$
\cite{Stone2008}, Li$_2$VO(Si,Ge)O$_4$ \cite{Melzi2000}, and
SrCu$_2$(BO$_3$)$_2$ \cite{Kageyama1999}, respectively.

In {\it unfrustrated} antiferromagnetic Heisenberg bilayers, QPTs have been
analyzed in great detail \cite{Wang2006}. Three-dimensional O(3)
universality and the critical coupling ratios have been firmly
established. For {\it frustrated} interdimer exchange, less research
has been performed.  In this context, very recently, bilayers of the Kitaev
quantum magnet on the ho\-ney\-comb lattice, coupled by interlayer
Heisenberg exchange, have attracted significant attention
\cite{Tomishige2018, Seifert2018, Tomishige2019, Koga2019}.  A prime reason
for this is, that in the limit of weak interlayer coupling, and instead of a
local magnetic order parameter of Ginzburg-Landau type, the single layers of
this magnet display a quantum spin liquid (QSL) state.

Indeed, the Kitaev single-layer magnet (KSLM) on the honeycomb lattice is one of
the few spin models, in which a $\mathbb{Z}_{2}$ QSL can exactly be shown to
exist \cite{Kitaev2006}. The spin degrees of freedom of this model fractionalize
in terms of mobile Majorana fermions coupled to a static $\mathbb{Z}_{2}$ gauge
field \cite{Kitaev2006, Feng2007, Chen2008, Nussinov2009, Mandal2012}.  In
finite external magnetic fields, chiral Majorana edge-modes arise. Transition
metal compounds with local Kramers doublets, induced by strong spin-orbit
coupling may serve to realize the KSLM \cite{Jackeli2009, Chaloupka2010}, with
$\alpha$-RuCl$_{3}$ \cite{Plumb2014} being among the most promising
candidates. However, due to ubiquitous additional non-Kitaev interlayer
exchange, only proximate Kitaev-QSLs have been reported so far
\cite{Banerjee2016, Banerjee2017, Banerjee2018, Trebst2017, Winter2017}. 
Remarkable advances to engineer the KSLM in ultra cold gases, trapped in 
optical lattices have also been made \cite{Duan2003}.

Direct realizations of a 
Kitaev-Heisenberg bilayer magnet (KHBM) are yet lacking. However, excitonic 
magnetism in van Vleck $d^4$ Mott insulators, like Li$_2$RuO$_3$ 
\cite{Miura2007} and Ag$_3$LiRu$_2$O$_6$ \cite{Kimber2010}, has been suggested 
to effectively model very closely related physics \cite{Khaliullin2013, 
Anisimov2019, Chaloupka2019}.

The genuine KHBM \cite{Tomishige2018, Seifert2018, Tomishige2019, Koga2019}
displays a remarkably rich set of quantum phases, which depend decisively on
the three different ways to stack the bilayer, the intralayer anisotropy of
the Kitaev exchange, as well as the ratio of the dimer- to Kitaev-exchange
\cite{Seifert2018}. In particular, and apart from the limiting cases of the
weakly interacting quantum dimer phase (QDM) and the Kitaev QSL, an emergent
Ising macro-spin phase as well as phase with spontaneous interlayer coherence
and flux generation has been uncovered within the phase diagram
\cite{Seifert2018}.

Apart for the quantum phases, the elementary excitations in dimer magnets are
of equal interest. This includes the dispersion \cite{Rueegg2005, Plumb2016}
and transport properties \cite{Romhanyi2015} of triplet excitations
(triplons) in the QDM, as well as the fate of Goldstone and Higgs modes on
approaching QPTs out of magnetically ordered phases \cite{Merchant2014,
Rueegg2008}. Apart from single triplon states, dimer magnets also allow to
study {\em multi triplon excitations} including the formation of triplon
(anti)bound states (ABS) \cite{Knetter2000, Collins2008, McClarty2017}, which
may be observed in spectroscopies, like ESR \cite{Nojiri1999}, Raman
\cite{Lemmens2000}, and inelastic neutron scattering \cite{Aso2005}.

For the KHBM some insight has been gained into the one-triplon dynamics in
its QDM phase in Ref. \cite{Seifert2018}. An analysis of the two-triplon
excitations and their interactions however is lacking. Therefore, in this
work we step forward and study the two-triplon spectrum and the emergence of
ABS in the KHBM. Moreover we consider the impact of these excitations on an
experimental observable, i.e., on magnetic Raman spectroscopy. The outline of
the paper is as follows. In Sec. \ref{sec:model} we detail the model. In
Sec. \ref{sec:method} we describe our method of calculation, i.e., the
perturbative Continuous Unitary Transformation (pCUT) and explain its
application to the evaluation of one- and two-triplon excitations. Results
for the one- and two-triplon spectra are presented in Sec. \ref{sec:twopart},
separately for the isotropic (Sec. \ref{sec:twopart:iso}) and anisotropic
case (Sec. \ref{sec:twopart:aniso}). The calculation of dynamical
correlation functions of observables and our findings for the Raman response
are described in Sec. \ref{sec:raman}. We conclude in
Sec. \ref{sec:conclusion}.  Several technical aspects are deferred to
appendices \ref{sec:appendix:spectrum}, \ref{sec:appendix:cluster} and
\ref{sec:appendix:raman}.

\begin{figure}[tb]
\includegraphics[width=0.7\columnwidth]{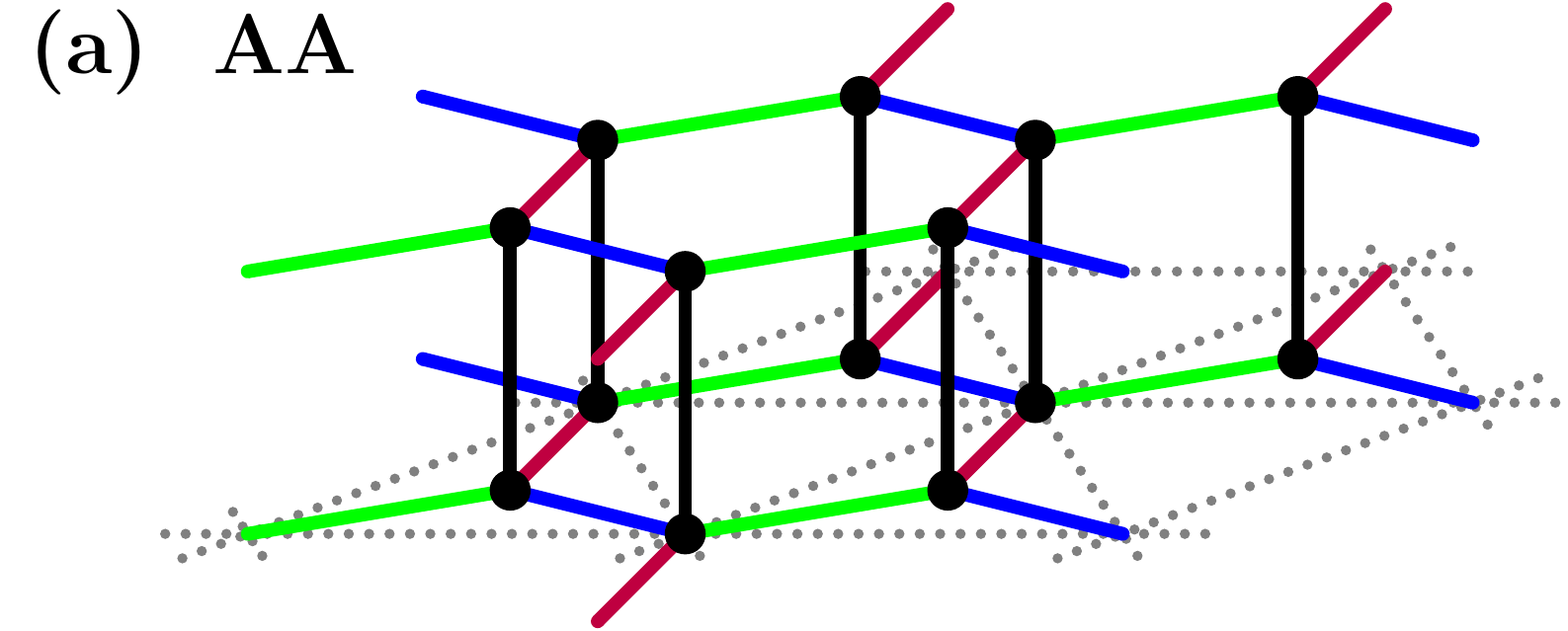}
\includegraphics[width=0.7\columnwidth]{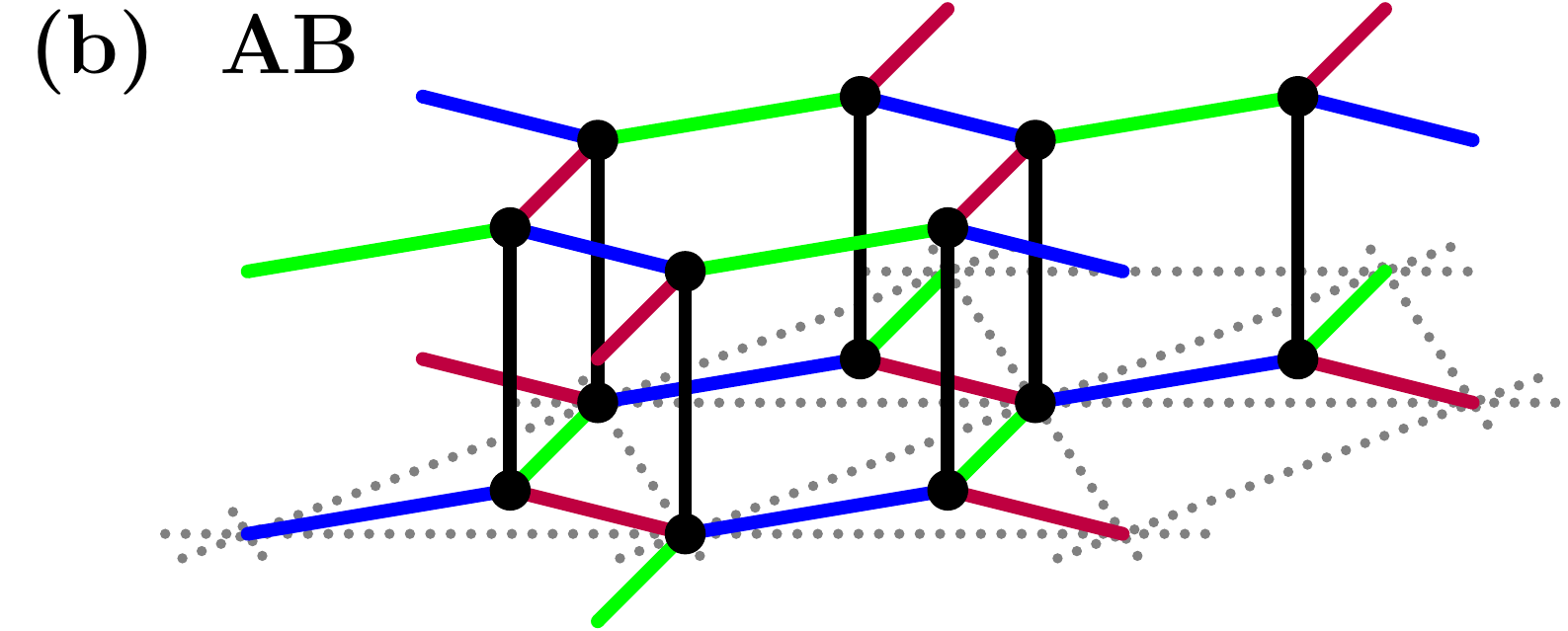}
\includegraphics[width=0.7\columnwidth]{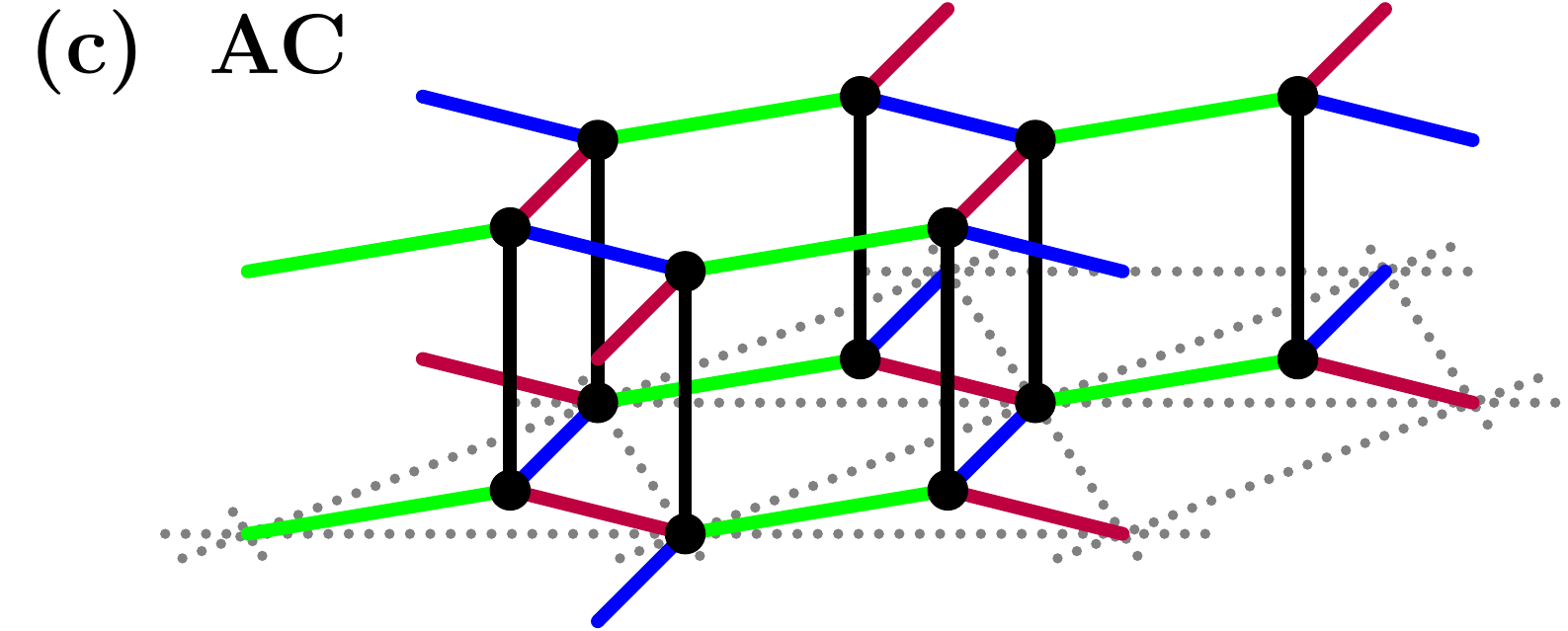}
\caption{The KHBM lattice. Solid dots: spin-$1/2$ operators. Solid green, blue,
red, and black bonds: $xx$, $yy$, $zz$ Ising, and Heisenberg exchange,
$J_\alpha$, and $J_\perp$, respectively. (a), (b), and (c): AA, AB, and AC
stacking.}
\label{fig:1}
\end{figure}

\section{Model}
\label{sec:model}

The Hamiltonian of the KHBM reads
\begin{equation}
\begin{aligned}
& H    =  H_0 + H_I \\
& H_0  =  J_\perp \sum_{\br_{j=1,2}} \vec{S}_{\br_j,1}\cdot\vec{S}_{\br_j,2} \\
& H_I = \sum_{\mathbf{r},\alpha,L} J_\alpha
S^\alpha_{\mathbf{r},L}
S^\alpha_{\mathbf{r}+\boldsymbol{\delta}_{\alpha}^L,L}
\end{aligned}
\label{eq:1}
\end{equation}
where $\vec{S}=\{S^\alpha\}$ with $\alpha=x,y,z$ are spin-1/2 operators,
$J_\perp$ and $J_\alpha$ are the Heisenberg interlayer and Kitaev intralayer
exchange, respectively. In this work we focus on antiferromagnetic 
(ferromagnetic) interlayer (intralayer) exchange $J_\perp > 0$ ($J_\alpha < 0$).
$L=1,2$ labels the two layers, with ${\bf r}=n_{1}{\bf
R}_{1}+n_{2}{\bf R}_{2}$ running over the sites of the triangular lattice where
${\bf R}_{1[2]}=(1,0), \,[(\frac{1}{2}, \frac{\sqrt{3}}{2})]$, with
$n_{1,2}\in\mathbb{Z}$, and $\boldsymbol{ \delta }^1_\alpha =
(\frac{1}{2},\frac{1}{2\sqrt{3}}),$ $(-\frac{1}{2}, \frac{1}{ 2\sqrt{3}})$,
$(0,-\frac{1}{\sqrt{3}})$ refer to the honeycomb basis sites of layer $1$,
tricoordinated to each bravais lattice site. $\br_{j=1,2} =
(\mathbf{r},\mathbf{r}+\boldsymbol{ \delta}^1_x )$ encodes the basis
of the honeycomb lattice, with ${\bf 0}_j = ({\bf 0}, \boldsymbol{ 
\delta}^1_x)$.

The model allows various {\it stackings}, by setting $\boldsymbol{ \delta}^2_{
\alpha}$ to certain permutations of $\boldsymbol{ \delta}^1_\alpha$ with
respect to $\alpha$.  More specifically, and as shown in
Fig. \ref{fig:1}, the so-called AA-stacking results from $\boldsymbol{
\delta}^2_\alpha=\boldsymbol{ \delta}^1_\alpha$, AB-stacking is obtained from
$\boldsymbol{ \delta }^2_{x,y,z}=\boldsymbol{ \delta}^1_{z,x,y}$, i.e., a 120°
rotation, and finally AC-stacking refers to $\boldsymbol{ \delta
}^2_{x,y,z}=\boldsymbol{ \delta}^1_{x,z,y}$, i.e., a mirror refection along one
of the Ising bonds. For anisotropic Ising exchange, i.e., $J_\alpha\neq
J_{\alpha'}$, it may be preferable to further sub-categorize these stackings
\cite{Seifert2018}.

\section{Method}
\label{sec:method}

The KHBM comprises at least two limiting quantum phases which are adiabatically
disjoint, i.e., are a separated by a QPT. For $|J_\alpha / J_\perp|\ll 1$ a QDM 
of weakly interacting antiferromagnetic dimers prevails. The dynamics of its
elementary triplon excitations strongly depends on the stacking. In particular
in the AA stacking the one-triplon excitations remain dispersionless for all
values of $J_\alpha / J_\perp$ due to an extensive number conserved quantities
build from the $\mathbb{Z}_2$ plaquettes of the two KSLMs, effectively locking 
the triplons on NN dimer pairs \cite{Tomishige2018, Seifert2018}. For 
$|J_\alpha / J_\perp|\gg 1$, two decoupled Kitaev $\mathbb{Z}_2$ QSLs occur 
with Majorana fermion and vison excitations. Between these, and at intermediate 
coupling $|J_\alpha / J_\perp|\sim 1$ additional quantum phases have been 
discovered \cite{Seifert2018}. Our prime goal for the remainder of this work is 
to stay within the QDM and focus on its two-triplon excitations, their 
interactions, the potential formation of bound states and possible consequences 
for observable spectroscopic probes.

The limit $|J_\alpha / J_\perp|\ll 1$ is amenable to {\it analytic} series
expansions (SE) by the perturbative Continuous Unitary Transformation (pCUT),
which is our method of choice. The prime ingredient of this method is, that
$H_0$ has a non-degenerate ground state and an equidistant ladder spectrum with
interlevel spacing $J_\perp$, the energy levels of which carry a monotonously
increasing integer quantum, or so-called {\it particle} number $Q$. For the
present $H_0$, $Q$ refers to the number of excited triplons, which for each
dimer we set to be
\begin{equation}
\begin{alignedat}{2}
&\ket{t_x} = -&(\ket{\uparrow\uparrow}-\ket{\downarrow\downarrow}) /
\sqrt{2} \,, \\
&\ket{t_y} = i&(\ket{\uparrow\uparrow}+\ket{\downarrow\downarrow}) / 
\sqrt{2} \,, \\
&\ket{t_z} =  &(\ket{\uparrow\downarrow}+\ket{\downarrow\uparrow}) /
\sqrt{2} \,,
\end{alignedat}
\label{eq:2}
\end{equation}
and refer to them as $x$-, $y$- and $z$-triplons hereafter. $Q=0$ refers to the
product ground-state $| \rangle = \prod_{\br_l}| s_{\br_l}\rangle$ of singlets. 
One- and two-triplon states, i.e., $Q=1$ and $2$, are $|\br_l\alpha \rangle = | 
t_{\br_l\alpha }\rangle \otimes \prod_{ \br_j \neq \br_l} |s_{ \br_j } \rangle$ 
and $|\br_j, \br_l, \alpha \beta \rangle = | t_{\br_j \alpha } t_{ \br_l \beta 
} \rangle \otimes \prod_{ \br_n \neq \br_j, \br_l} |s_{ \br_n } \rangle$.
In general the perturbation $H_I$ mixes different Q-sectors, i.e., the
block-diagonal form of $H_0$ with respect to the triplon number is not
conserved for $H$. Based on the ladder spectrum of $H_0$ however, a unitarily
equivalent effective Hamiltonian $H_\mathrm{eff}=U^\dagger H U$ can be
constructed which remains $Q$-{\it diagonal}, using the flow equation method of
Wegner \cite{Wegner1994}. This methods provides the link to SE,
as it is implemented perturbatively order by order in $H_I$, leading to
\begin{equation} 
H_{\mathrm{eff}}=H_0 + \sum_{l,m,n}^\infty C_{l,m,n} J_x^l
J_y^m J_z^n, \label{eq:3} 
\end{equation} 
where $C_{l,m,n}$ are weighted products
of terms in $H_I$, comprising $l+m+n$ non-local creations(destructions) of
triplons which conserve the $Q$-number. The weights are determined by recursive
differential equations. Details can be found in Refs. \cite{Knetter2000a, 
Knetter2003, Knetter2003t}.

$Q$-number conservation allows to obtain the spectrum of $H$ from the
irreducible matrix elements of $H_{\rm eff}$ with respect to the eigenstates of
$H_0$. The ground state, i.e., $Q=0$, energy is $E_0=\langle | H_{\rm eff} |
\rangle$. The one-triplon, i.e., $Q=1$, dispersion results from ${\bf 
E}_{\mathbf{k},jl,\alpha\beta} = \sum_{\bf r} e^{i\br_j \cdot {\bf k} } \langle 
\br_j\alpha | H_{\rm eff} | {\bf 0}_l \beta \rangle - \delta_{ \br_j {\bf 0}_l} 
\delta_{\alpha\beta} E_0^{\text{cl}}$. In view of the two-site basis $j=1,2$ 
and the spin index $\alpha$, this is a $6 \times 6$-matrix. However, as 
discussed in appendix \ref{sec:appendix:spectrum}, the one-triplon hopping 
matrix elements are diagonal regarding the triplon components, leaving three $2 
\times 2$ dispersions, with roots $E_{\mathbf{k},i=1,2,\alpha}$, which can be 
determined analytically \cite{Seifert2018}.

\begin{figure}[tb]
\includegraphics[width=0.49\columnwidth]{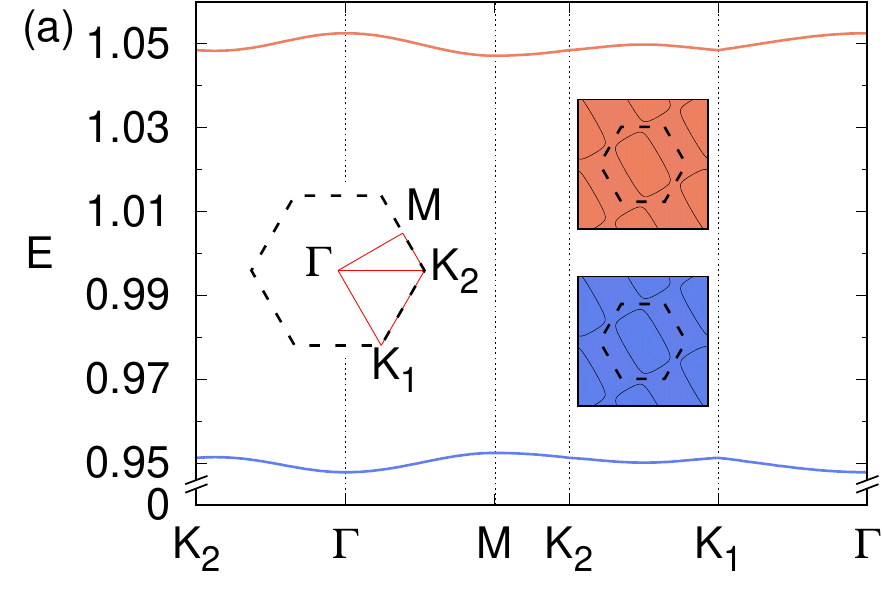}
\includegraphics[width=0.49\columnwidth]{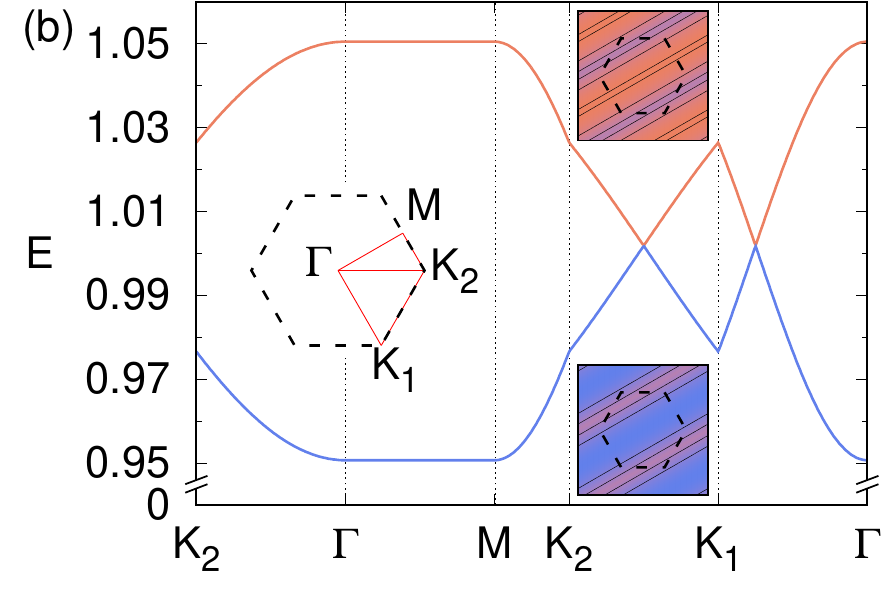}
\caption{
Solid blue(orange) line: one-triplon dispersion $E_{\mathbf{k},i=1(2),\alpha}$
for (a) $\alpha=x$-triplons and (b) $\alpha=y[z]$-triplons in AC-stacking at $J
= -0.1 J_\perp$ along high-symmetry lines of BZ. Insets: constant energy
surfaces.
}
\label{fig:2}
\end{figure}

The $Q=2$, i.e., the two-triplon problem is more challenging \cite{Knetter2003, 
Knetter2003t}. First, the irreducible two-triplon interaction is evaluated. 
Primarily this requires calculation of the matrix elements $\langle \br'_{j'}, 
\br'_{j'}+ \boldsymbol{\tau}'_{l'}, \alpha' \beta'| H_{\rm eff} | \br_j, \br_j+
\boldsymbol{\tau}_l, \alpha \beta \rangle$ for initial (final) two-triplon
positions $\br^{(\prime)}_{j^{(\prime)}}, \br^{(\prime)}_{j^{(\prime)}}+
\boldsymbol{\tau}^{(\prime)}_{l^{(\prime)}}$ and triplon components
$\alpha^{(\prime)} \beta^{(\prime)}$. Again, the subscripts $j,l=1,2$ label the
honeycomb basis. Second, using translational invariance, a two-triplon
Hamiltonian {\it matrix} $h_{\bf K}( \boldsymbol{ \tau}^\prime_{ l^\prime},
\boldsymbol{ \tau}_{l}, \alpha' \beta', \alpha \beta)$ is set up from this,
between two-triplon states $| {\bf K}, \boldsymbol{\tau}_l,\alpha \beta
\rangle$, classified according to total momentum ${\bf K}$, two-triplon
separation $\boldsymbol{\tau}_l$, and spin components $\alpha \beta$. Third,
$h_{\bf K}( \boldsymbol{ \tau}^\prime_{ l^\prime}, \boldsymbol{\tau}_{l},
\alpha' \beta', \alpha \beta)$, comprising {\it analytical} matrix elements from
the SE, is diagonalized {\it numerically} on sufficiently large systems.

In this context it is essential to understand the physics of $h_{\bf K}(
\boldsymbol{ \tau}^\prime_{ l^\prime}, \boldsymbol{\tau}_{l}, \alpha' \beta',
\alpha \beta)$. Namely, since it describes a two-body scattering problem for
each ${\bf K}$, it is a banded matrix regarding $\boldsymbol{ \tau}^\prime_{
l^\prime}, \boldsymbol{\tau}_{l}$. I.e., only in its upper left corner, for
two-triplon separations $|\boldsymbol{ \tau}^{(\prime)}_{ l^{(\prime)}}| < d_I$,
less than a characteristic distance, actual interactions occur. On the remainder
of the band diagonal, free scattering states reside. This knowledge directly
allows to eliminate any finite size effects of the numerical solution of this
two-body scattering problem.  We emphasize, that the systems involved in
this final numerical step of the pCUT procedure easily comprise more than
several hundred sites.

Similar to the one-triplon case, simplifications apply to the two-triplon
Hamiltonian regarding its $9\times 9$ spin structure. In fact, non-zero
irreducible matrix elements arise only for $\alpha'\beta' = \alpha\beta$ with
$\alpha\beta$ one out of $(x,y)$, $(y,z)$ and $(z,x)$, or within the $3\times 3$
block of $\alpha^{(\prime)} \beta^{(\prime)} = [(x,x)$, $(y,y)$, $(z, z)]$. For
additional details on the $Q=2$ case, see appendix \ref{sec:appendix:spectrum}.

All evaluations of matrix elements of $H_{\rm eff}$, referred to in the
preceding, are carried out on suitably chosen linked cluster graphs of the
lattice. A description of this is deferred to appendix 
\ref{sec:appendix:cluster}.

\section{One- and Two-triplon excitations}
\label{sec:twopart}

In this section we present our findings for the excitation spectra. As compared
to conventional bilayer quantum magnets and as a consequence of the anisotropic
Ising-type Kitaev exchange as well as the stacking options, we find these
spectra to have a remarkably rich structure. Our main focus is on the
two-triplon dynamic. However, and to determine the irreducible two-triplon
interactions, also the ground state energy and the one-triplon spectra of the
KHBM need to be known. These have been analyzed in Ref. \cite{Seifert2018}. For
clarity, we revisit selected aspects of the latter results first.

\subsection{One-triplon dispersion}
\label{sec:onepart}

Here we remain with the AC-stacking. Figs.  \ref{fig:2}(a) and (b) display the
dispersions $E_{\mathbf{k},i,\alpha}$ and the constant energy surfaces of the x-
and y-/z-triplons, at $O(9)$ respectively. Each figure contains two dispersions,
referring to the two-site basis. The plots evidences, that for a given type of
stacking, rather distinct one-triplon dispersions can arise, depending on the
type of triplon. This is a direct consequence of the directionally dependent
Kitaev exchange and does not only pertain to the kinetic energy, but also
induces an effective dimensionality. E.g., in the AC case, x-triplons acquire a
non-zero hopping matrix element only at $O(2)$ of the SE and remain almost
localized on the dimers linked by $S^x S^x$-exchange in Fig.  \ref{fig:1}(c). In
contrast, the y-/z-triplons disperse at $O(1)$, however, their hopping matrix
elements are locked to the zig-zag ladder in Fig.\ref{fig:1}(c), providing them
with an identical and effectively 1D dispersion. For additional properties of
the one-triplon excitations we refer to Ref. \cite{Seifert2018}.

One information, important for the two-triplon spectrum, which can be read off
directly from the one-triplon dispersion, is the support of the non-interacting
two-triplon continuum. Namely, at fixed total momentum ${\bf K}$ it refers to 
the range of all energies $[E_{\mathbf{K}-\mathbf{q},i,\alpha} +
E_{\mathbf{q},j,\beta}]$ with $\mathbf{q} \in \text{BZ}$.

\subsection{Isotropic Kitaev exchange} \label{sec:twopart:iso}

\begin{figure}[tb]
\mbox{\makebox[0cm]{(a)}\hspace{\columnwidth}\phantom{.}}
\\
\includegraphics[width=1\columnwidth,page=1]{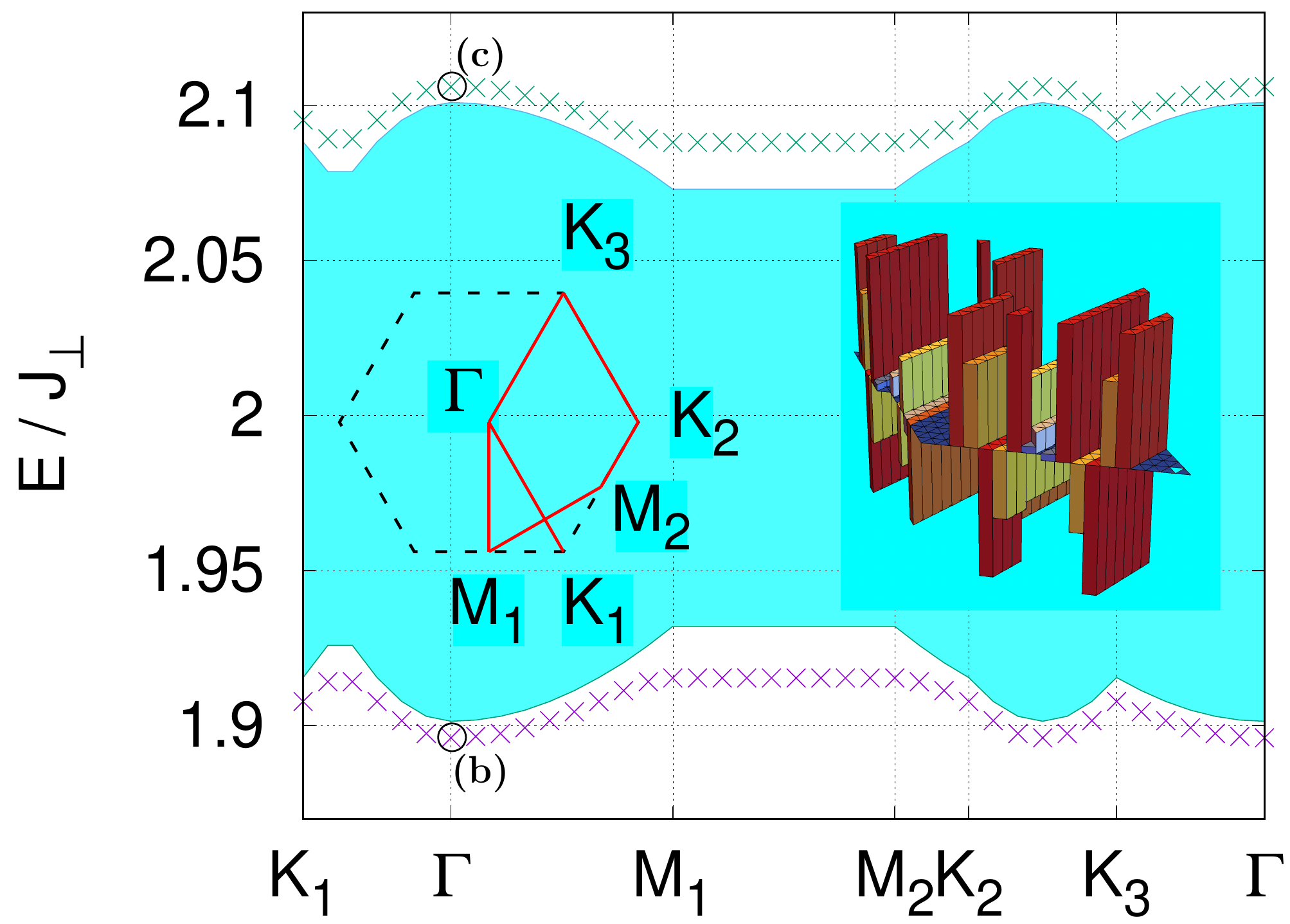}
\\
\mbox{\makebox[0cm]{(b)}\hspace{0.5\columnwidth}
\makebox[0cm]{(c)}\hspace{0.5\columnwidth}\phantom{.}}
\\
\includegraphics[width=0.49\columnwidth]{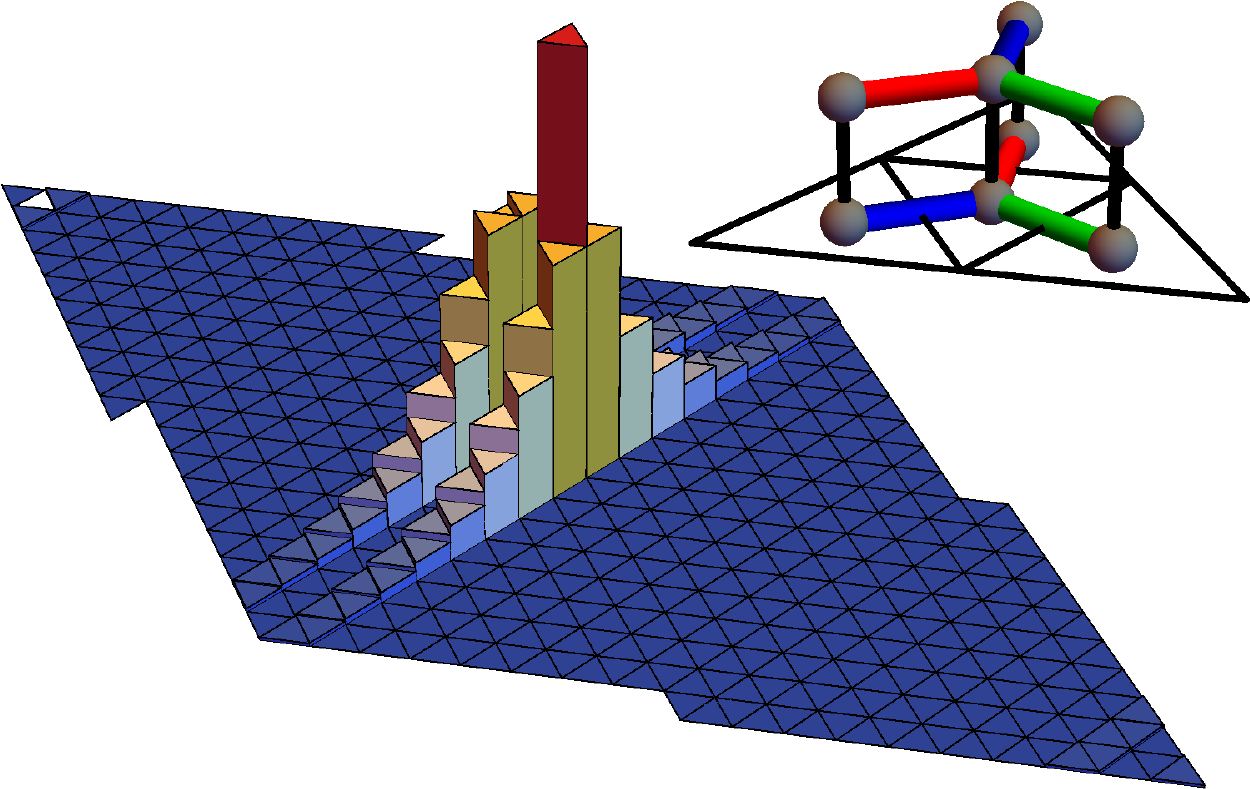}
\hfill
\includegraphics[width=0.49\columnwidth]{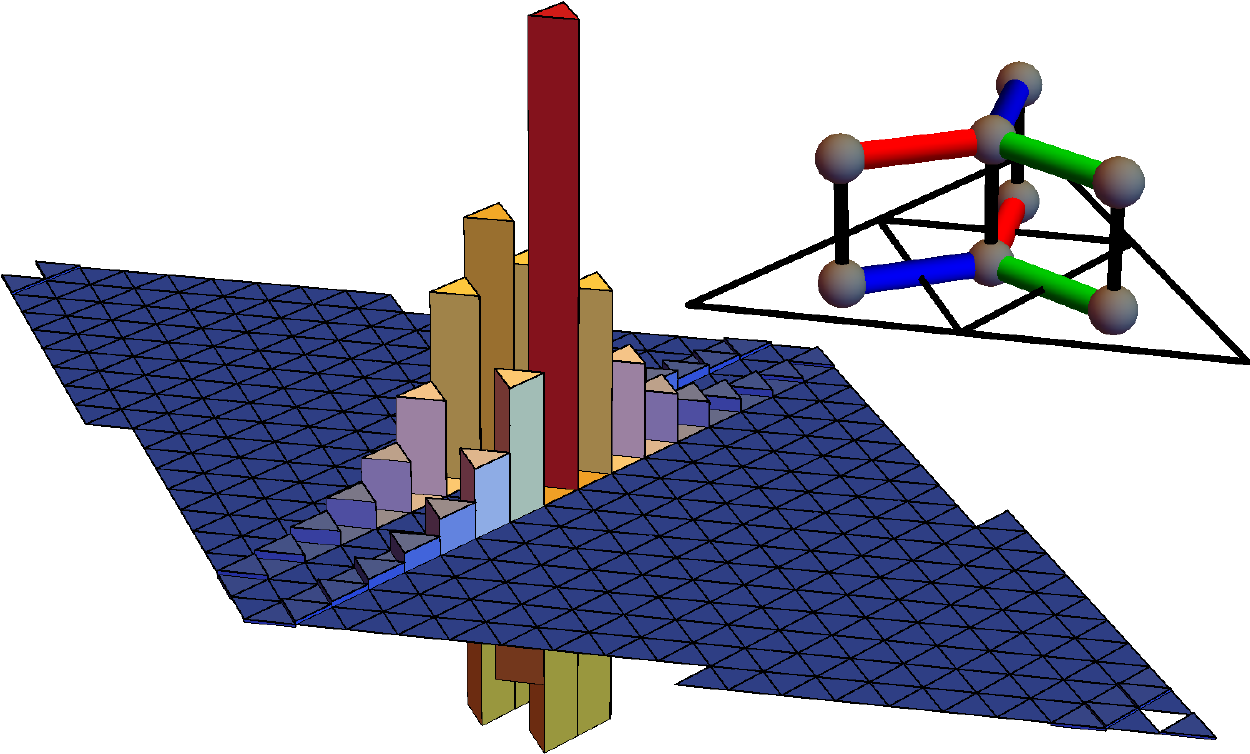}
\caption{
Panel (a): Two-triplon spectrum versus total momentum in $(y,z)$ channel along
high-symmetry path of BZ for AC-stacking at $J=-0.1 J_\perp$. Solid cyan region:
Scattering state continuum. (Green)Magenta crosses: (Anti)Bound states
(ABS). Panels (b) and (c): $yz$-component of wave function of ABS at total 
momentum ${\bf K}=\Gamma$, i.e., energies (b) and (c) in panel (a), versus real 
space coordinate $\boldsymbol{\tau}_l$ ($zy$-component provides identical 
information). Center of 2D plotting planes refer to $\boldsymbol{\tau}_l = {\bf 
0}_1$ and real-space bond-geometry is for reference. Analogous display within 
solid cyan region: Wave function of representative scattering state. System 
size: $18\times 18$ cells.
}
\label{fig:3}
\end{figure}

Now we turn to the two-triplon spectra, focusing on the isotropic case $J_{x, y,
z} = J$ first. We begin with the AC-stacking and discuss a central result of our
work in Fig. \ref{fig:3}. This displays the eigenenergies obtained from the
numerical diagonalization of the two-triplon subsector of the effective pCUT
Hamiltonian in the $(y,z)$ channel, described in Sec. \ref{sec:method}, and
obtained at $O(9)$ and $O(8)$ of the SE for one and two-triplon matrix elements,
respectively. The spectrum is shown along various high-symmetry directions for
the total two-triplon momentum in the BZ. Several comments are in order. First,
a broad continuum is visible, which refers to the energy range of the
two-triplon scattering states, comprising renormalized one-triplon dispersions
as in Fig. \ref{fig:2}. Second, there are clearly visible eigenenergies outside
of the continuum, both below and above. These are the sought-for {\it bonding}
and {\it antibonding} collective two-triplon states.  We find these ABS for all
coupling strengths within the convergence radius of the series and for all 
total momenta $\mathbf{K}$ investigated.

Third, and to prove the (anti)bonding nature of these states, we analyze not
only the energies of the two-triplon states, but also their wave functions
versus the relative inter triplon separation $\boldsymbol{\tau}_l$. One
representative case is shown in Fig. \ref{fig:3} for each of the bonding, the
antibonding, and the scattering-states spectral regions. The main difference
between the former two and the latter is, that the ABS show a strong 
localization versus $\boldsymbol{\tau}_l$, while the scattering states are 
rather evenly distributed, with at most short wave-length oscillations.

Finally we emphasize, that in stark contrast to conventional bilayer spin models
\cite{Knetter2000, Collins2008}, the KHBM's spectrum of ABS does not separate
into channels with well-defined total spin quantum-numbers, because of its
broken $SU(2)$ invariance.

The existence of the AC $(y,z)$ ABS in the KHBM allows for a
qualitative understanding. As noted in Sec. \ref{sec:onepart}, the y- and
z-triplons in the AC-stacking spread essentially in 1D along the $S^{y(z)}
S^{y(z)}$ zigzag chains. One may now envisage an (anti)symmetric linear
combination $\ket{\Psi_\pm}$ of the corresponding Wannier states $\ket{x}_a$,
$\ket{y}_b$ on two dimers, linked by $S^x_{a,L} S^x_{b,L}$ intralayer exchange
bonds
\begin{equation}
\ket{\Psi_\pm} = \ket{y,z}_{ab} \pm \ket{z,y}_{ab}\,.
\label{eq:4}
\end{equation}
These are eigenstates of the local intralayer 
$x$-exchange, combined between both layers
\begin{equation}
J_x ( \sum_{L=1,2} S^x_{a,L} S^x_{b,L} ) \ket{\Psi_\pm} = \mp
\frac{1}{2} J_x
\ket{\Psi_\pm}
\label{eq:5}
\end{equation}
This lowest order on-bond (repulsion)attraction by
$(+)-J$/2 is the reason for the formation of the ABS. As our results show this
general relation holds true even for higher order SE. The preceding is
corroborated by the two figures regarding the (anti)bound wavefunctions in
Fig. \ref{fig:3}(b) and \ref{fig:3}(c). In fact, their probability amplitudes 
are confined to both triplons being on neighboring y/z-zigzag chains coupled by
intralayer $S^x S^x$ bonds. As to be expected the wavefunction amplitudes get
monotonously smaller for larger triplon distances because higher order coupling
terms are needed to introduce two-triplon interactions at larger distances.

\begin{figure}[tb]
\mbox{\makebox[0cm]{(a)}\hspace{\columnwidth}\phantom{.}}
\\
\includegraphics[width=1\columnwidth,page=1]{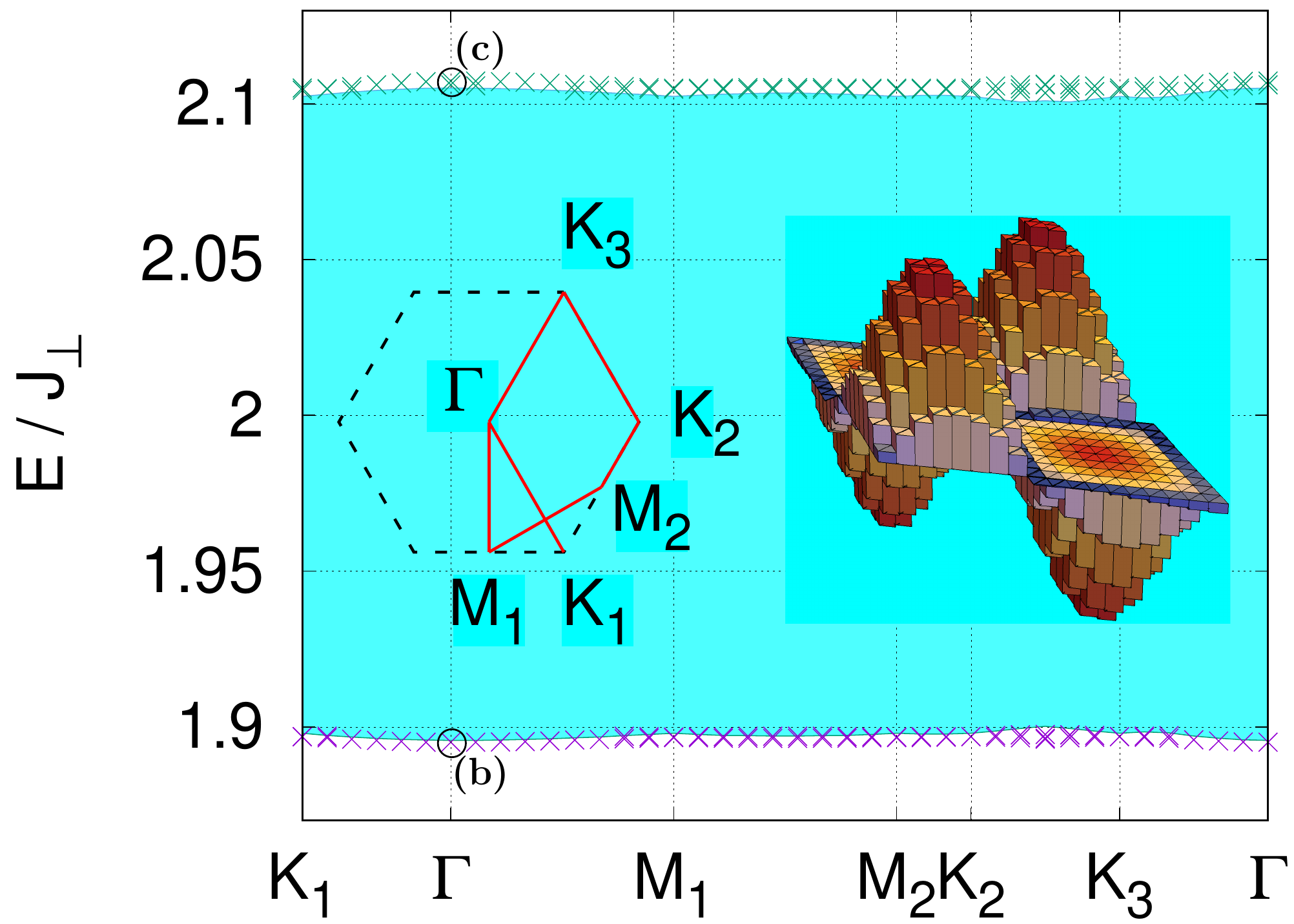}
\\
\mbox{\makebox[0cm]{(b)}\hspace{0.5\columnwidth}
\makebox[0cm]{(c)}\hspace{0.5\columnwidth}\phantom{.}}
\\
\includegraphics[width=0.49\columnwidth]{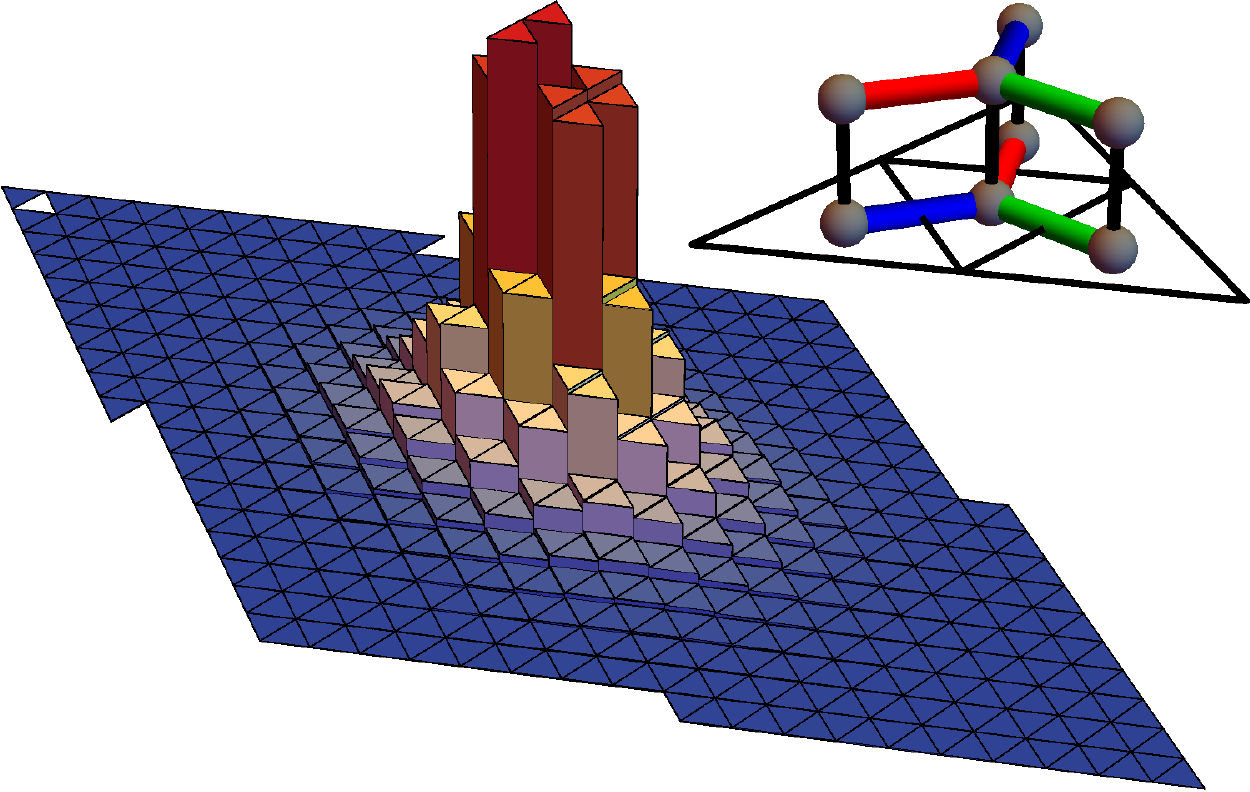}
\hfill
\includegraphics[width=0.49\columnwidth]{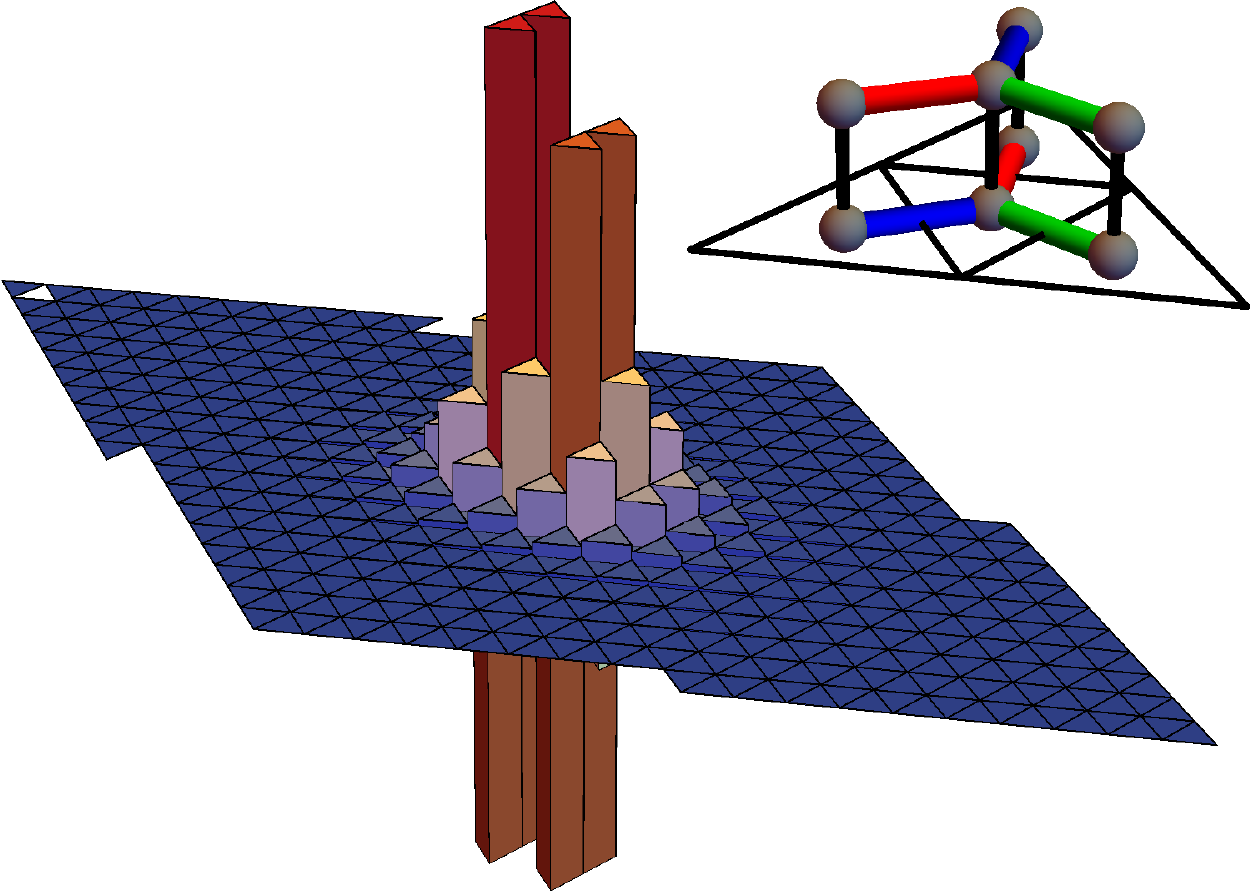}
\caption{
Panel (a): Two-triplon spectrum versus total momentum in the $(x,x), (y,y),
(z,z)$-sector along high-symmetry path of BZ for AC-stacking at $J=-0.1
J_\perp$. Solid cyan region: Scattering state continuum. {\em Three} sets of
(green)magenta crosses: (Anti)Bound states (ABS). Panels (b) and (c): 
$xx$-component of wave function of ABS at total momentum ${\bf K}=\Gamma$, 
i.e., energies (b) and (c) in panel (a), versus real space coordinate 
$\boldsymbol{\tau}_l$. $yy$- and $zz$-components are negligibly small. Center 
of 2D plotting planes refer to $\boldsymbol{\tau}_l = {\bf 0}_1$ and real-space 
bond-geometry is
for reference. Analogous display within solid cyan region: Wave function of
representative scattering state. System size: $18\times 18$ cells.
}
\label{fig:4}
\end{figure}

Now we turn to the $(x,x),(y,y),(z,z)$-sector. There, ABS similar to Eqs. 
(\ref{eq:4},\ref{eq:5}) can be constructed, namely
\begin{equation}
\ket{\widetilde{\Psi}_\pm} = \ket{y,y}_{ab} \pm \ket{z,z}_{ab}\,.
\label{eq:6}
\end{equation}
These are eigenstates of the local intralayer $x$-exchange, analogous to
Eq. (\ref{eq:5}), with on-bond (repulsion)attraction, and
therefore lead to ABS of identical origin. Their splitting off
the continuum however, is much less than in the $(y,z)$ sector because the
contribution from the $(x,x)$ triplons, see Fig. \ref{fig:2},
broadens the continuum of the $(x,x),(y,y),(z,z)$-sector, which reduces the
two-triplon binding energies. The corresponding complete spectrum is shown in 
Fig. \ref{fig:4}. Indeed, we find that these ABS are observable only for small 
$J$ and only in some regions in the BZ. They also remain much closer to the 
continuum than the $(yz)$ states. On the contrary we find a different type of 
ABS which is build around $\ket{x,x}$ states. Representations of these ABS are 
shown in Fig. \ref{fig:4}(b) and \ref{fig:4}(c). The states have a 
non-one-dimensional structure other than the $(yz)$ states but are similarly 
localized. For $J \gtrsim 0.3$ all ABS merge with the continuum for most of the 
BZ, except for the antibound states in the vicinity of the line from $K_2$ to 
$K_3$ which stay above the continuum.

All remaining spin-$\alpha\beta$ two-triplon states display no irreducible
interaction and therefore only free two-triplon continua with no collective
states. 

To summarize, the two-triplon problem has four distinct channels. In the 
$(y,z)$-channel we find a single type of ABS, while in the 
$(x,x)(y,y)(z,z)$-channel we find two distinct types of ABS. The two remaining 
channels show no ABS because the $x$-triplon does not form an ABS with $y$- or 
$z$-triplons.

Now we turn to the AA-stacking. As noted in Sec. \ref{sec:method}, here all
triplons remain dispersionless and confined to NN-dimer-pairs
\cite{Tomishige2018, Seifert2018}. Therefore this case is very exceptional in as
such that dispersive two-triplon scattering states and the corresponding
spectral continua are absent. However, two-triplon interactions, as described in
Eqs. (\ref{eq:4}) and (\ref{eq:6}) are still active, leading to completely
localized ABS whenever two triplons are placed close enough. Therefore, the
complete spectrum on an $O(N\times N)$ system comprises several discrete levels,
some of which are $O(N^4)$ degenerate ('scattering states') and some of which
are $O(N^2)$ degenerate ('ABS').

\begin{figure}[tb]
\includegraphics[width=0.925\columnwidth]{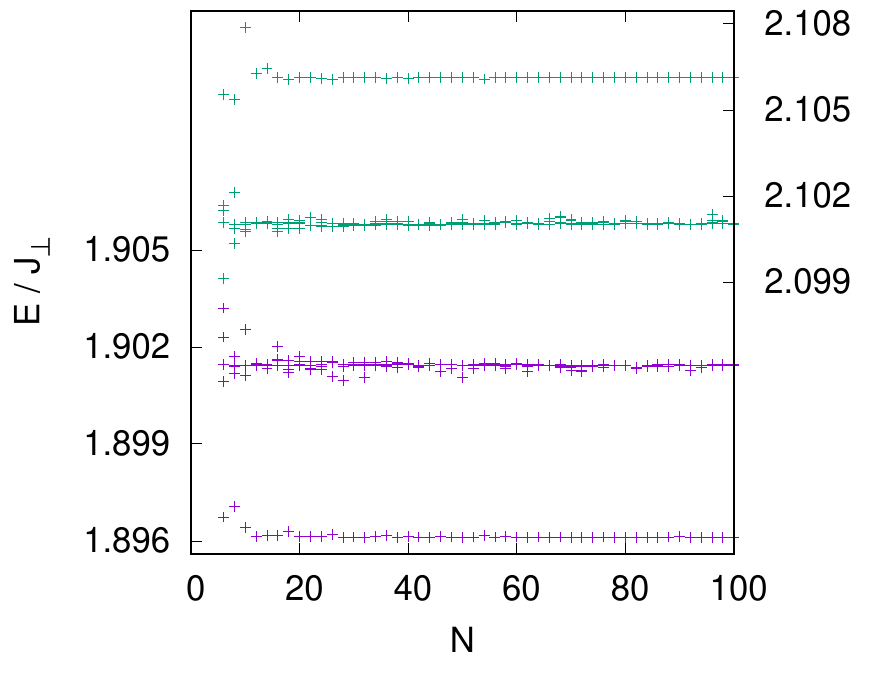}
\caption{Left(right) y-axis: Two lowest(highest) eigenvalues, i.e., ABS and
continuum boundary for each, of two-triplon spectrum versus linear size of
$N\times N$ unit-cell system in $(y,z)$-sector of AC-stacking with $J_\alpha =
-0.1 J_\perp$ at zero total momentum, $\mathbf{K} = \mathbf{0}$. Finite size
effects are negligible beyond $N\gtrsim 15$.}
\label{fig:5}
\end{figure}

As for the last remaining type of stacking, i.e., AB, we find no clear 
signatures of ABS at the maximum order of the SE we have generated and for 
those intralayer couplings acceptable regarding the convergence of the SE. In 
fact, we have observed weak indications of ABS for this stacking only in 
limited regions of the BZ and only for $J\gtrsim 1.1$. This we consider unsafe 
regarding the SE. Interestingly, the absence of ABS for the AB-stacking is 
consistent with the fact, that local two-triplon eigenstates analogous to Eqs. 
(\ref{eq:4}) and (\ref{eq:6}) cannot be constructed in this case, because two 
dimers are never coupled by identical Ising exchange from both Kitaev layers.

Finally, we comment on the control of finite size effects in the diagonalization
of the $Q=2$ sector of the effective Hamiltonian. The flow of several low- and 
high-energy eigenstates with system size is shown in Fig.  \ref{fig:5}. As
can be seen, errors due to finite lattice size can be discarded beyond
$N\simeq 15$. This is consistent with the strong localization of the ABS. In 
practice we have employed $N=18$.

\subsection{Anisotropic Kitaev exchange}
\label{sec:twopart:aniso}

For anisotropic Kitaev exchange, i.e., $J_\alpha/J_\beta\neq 1$ for $\alpha \neq
\beta$ in general, we focus on the AC-stacking at $J_x \neq J_y = J_z$. This
case is of interest, as it provides additional insight into the qualitative
picture for the occurrence of the ABS provided in the context of
Eqs. (\ref{eq:4}) and (\ref{eq:5}).

As described in Secs. \ref{sec:onepart} and \ref{sec:twopart:iso}, for
AC-stacking, the one-triplon hopping is quasi one-dimensional along the $yz$
zigzag ladder and the width of the one-triplon dispersion, as well as that of
the two-triplon scattering-state continuum is therefore primarily determined by
the size of $J_{y,z}/J_\perp$. Moreover, to lowest order, i.e., for $J_{y,z} = 
0$, the ABS (anti)binding energy is set by $J_x$ from Eqs. (\ref{eq:4}) and 
(\ref{eq:5}).

\begin{figure}[tb]
\includegraphics[width=1\columnwidth,page=2]{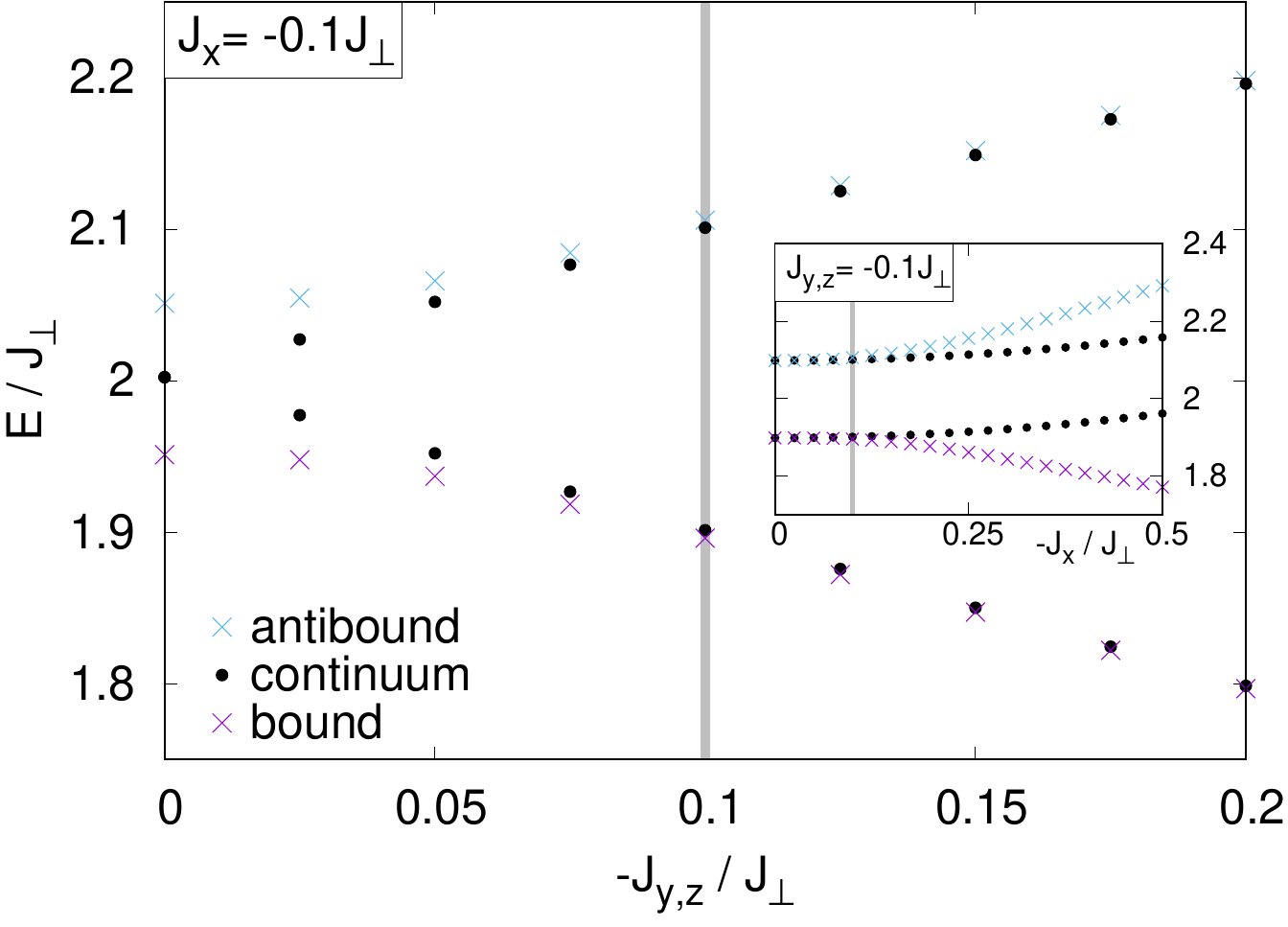}
\caption{Cyan/magenta crosses: ABS energies and solid dots: upper/lower
scattering-state continuum bounds versus $J_{y,z}$ at $J_y=J_z$ for fixed $J_x =
-0.1 J_\perp$ in the  $(y,z)$ channel. Gray vertical line: $J_x = J_y = J_z$. 
Inset: Identical quantities, however versus $J_x$ for fixed $J_{y,z} = -0.1 
J_\perp$.}
\label{fig:6}
\end{figure}

Performing pCUT at $J_x \neq J_y = J_z$ therefore allows to asses the interplay
between the binding and the dispersion. This is shown in Fig. \ref{fig:6}. 
First, at fixed $J_x$, varying $J_{y,z}$ one can clearly see that, while for
$J_{y,z}=0$ the ABS binding is strongest {\em and} exactly of the magnitude
given in Eq. (\ref{eq:5}), finite triplon dispersion 
reduces the splitting of the ABS off from the scattering-state continuum.  
However, even for $J_{y,z} \gtrsim J_x$ some, albeit small (anti)binding energy
remains. Similarly, we can fix $J_{y,z}\neq 0$, as in the inset of Fig. 
\ref{fig:6} and increase $J_x$ beyond $J_{y,z}$. This enhances the binding
as compared to the dispersion and hence increases the split-off of the ABS from
the continuum. In summary, we observe an adiabatic evolution of the ABS versus
$J_x / J_{y,z}$, which strongly supports the simple physical picture set forth
in Eq. (\ref{eq:4}).

\section{Raman spectrum}
\label{sec:raman}

Having discussed the energies of the $Q=1$ and 2 excitations, we now turn to
observables. While one-triplon states can be probed by neutrons, Raman
scattering is a prime tool to analyze two-triplon excitations in quantum magnets
in general \cite{Lemmens99}, including, e.g., paradigmatic frustrated systems
\cite{perkins08}, Kitaev magnets \cite{Wulferding2020}, and dimer magnets
\cite{Brenig2001, Jurecka2001}. For the latter, Raman scattering will also probe
the ABS. Therefore, our focus is on Raman scattering.

The Raman response is obtained from the spectrum of the dynamical correlation
function of the Raman operator, encoding the inelastic scattering of light from
the system. For local moment systems the latter is mediated by a photon-assisted
superexchange process, described by the Loudon-Fleury vertex \cite{Fleury68}
\begin{equation}
\mathcal{R} = \sum_{\mathbf{r},\alpha,L} S^\alpha_{\mathbf{r},L}
S^\alpha_{\mathbf{r}+\boldsymbol{\delta}_{\alpha}^L,L}
\left[\mathbf{A}^i\cdot\boldsymbol{\delta}_{\alpha}^L\right]\left[
\mathbf{A}^o\cdot\boldsymbol{\delta}_{\alpha}^L\right]
\label{eq:7}
\end{equation}

where $\mathbf{A}^{i,o}$ is the vector potential of the incoming and outgoing
light, using radiation gauge, $i\omega \mathbf{A}={\bf E}$. To simplify, we
confine ourselves to a scattering geometry in which the incoming and outgoing
polarization of the electric field is in the Kitaev plane. I.e., Eq. 
(\ref{eq:7}) has no contributions from the dimer
exchange. 

From $\mathcal{R}$ and the ground state $\ket{\psi}$, the zero-temperature
Raman intensity is given by
\begin{align}
- \pi I(\omega) & =  {\rm Im}
\bra{\psi}\mathcal{R}^\dagger \frac{1}{\omega+i\,0^+-
H+E_0}\mathcal{R}\ket{\psi} \nonumber \\
& = {\rm Im}
\bra{}\mathcal{R}_{\rm eff}^\dagger \frac{1}{\omega+i\,0^+-
H_{\rm eff}+E_0}\mathcal{R}_{\rm eff}\ket{}\,,
\label{eq:8}
\end{align}
where the second line refers to the application of the pCUT and, as for $H_{\rm
eff}$, the operator $\mathcal{R}_{\rm eff}$ is the pCUT transform of the Raman
operator, i.e., $U^\dagger \mathcal{R} U$ where $\ket{\psi}=U\ket{}$ by virtue 
of the pCUT \cite{Knetter2003t, Knetter2004}.  For the evaluation of 
$\mathcal{R}_{\rm eff}$, it is important to note, that pCUT provides the 
generator of the unitary transformation $U$ explicitely.

Since typically, $\mathbf{A}^{i,o}$ refers to light within the visible range, 
its wave length is exceedingly large compared to the lattice constant. 
Therefore, the Raman operator Eq. (\ref{eq:7}) generates excitations of
zero total momentum and Raman scattering probes only their spectral density. As
a practical consequence, only ${\bf K}=0$ sectors of $H_{\rm eff}$ are relevant
for $I(\omega)$.

While $H_{\rm eff}$ is $Q$-diagonal, $\mathcal{R}_{\rm eff}$ in general is
not. Therefore $I(\omega)$ constitutes a sum over spectra from subspaces of
various triplon numbers. We confine ourselves to the contribution comprising the
smallest number of triplons which occurs, namely $Q=2$. While this corresponds 
to the number of triplons created also by $\mathcal{R}$ itself, the $Q=2$ 
creation by $\mathcal{R}_{\rm eff}$ involves an operator of very different 
real-space extent than $\mathcal{R}$.

\begin{figure}[tb]
\includegraphics[width=1\columnwidth]{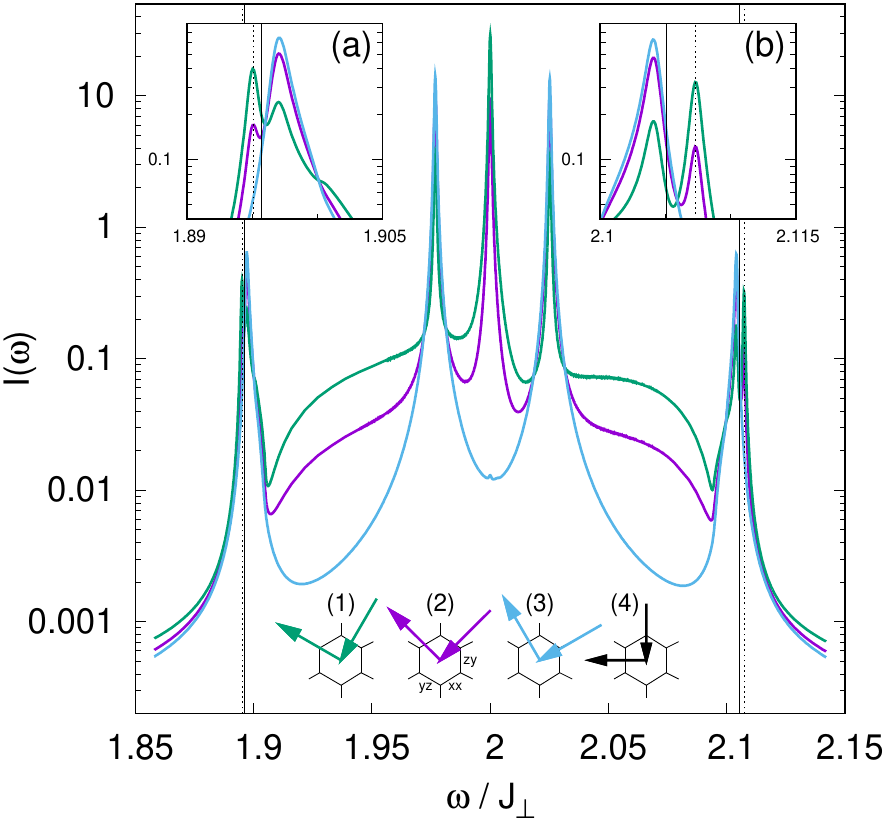}
\caption{Raman intensity $I(\omega)$ in $(x,x), (y,y), (z,z)$-sector of
AC-stacking at $J = -0.1 J_\perp$ (see Fig. \ref{fig:4}) versus energy transfer
$\omega$ for three distinct in-plane, perpendicularly polarized, scattering
geometries, shown by hexagons (1), (2) and (3). Due to the symmetry of the 
model $I(\omega)$ is identical for geometries (1) and (4). Letters 
$\alpha\beta$ on hexagon bonds label upper/lower layer $J_{\alpha/\beta}$ 
exchange. Dashed(solid) vertical lines: ABS(continuum boundary) energies. 
Insets (a) and (b): Blow up of intensity in vicinity of ABS.}
\label{fig:7}
\end{figure}

Based on these preliminaries, we evaluate $I(\omega)$ by continued fraction
expansion, using a Lanczos procedure for $H_{\rm eff}$ with the two-triplon
states generated by $\mathcal{R}_{\rm eff}\ket{}$ as the start vector. Details
can be found in appendix \ref{sec:appendix:raman} and in Ref. 
\cite{Knetter2003}.

Fig. \ref{fig:7} shows the Raman intensity in the AC-stacking for isotropic
Kitaev exchange $J_{x,y,z}=J$ calculated with one-triplon matrix elements to 
$O(9)$, two-triplon interactions to $O(8)$ and $\mathcal{R}_{\rm eff}\ket{}$ 
to $O(4)$. The imaginary shift off the real $\omega$-axis has been chosen small 
enough to resolve all intrinsic fine structure of the spectrum. For the 
truncation order of the continued fraction chosen, this implies some 
insignificant ripple, visible on the spectrum. For the scattering geometry we 
have chosen perpendicular polarization of the incoming and outgoing light, 
varying the angle relative to the tricoordinated nearest-neighbor bonds. 
Several points can be made. First, the spectrum displays up to five peaks, two 
of which, namely those on the edges of the continuum are down by approximately 
one order of magnitude in intensity. Second, the global shape of the intensity 
is strongly dependent on the scattering angle. In particular, the central peak 
at $\omega = 2$ and the contribution from the ABS - to be discussed shortly - 
can be suppressed completely for incoming polarization parallel to the stacked 
$S^xS^x$-bonds. In this configuration the Raman operator can only directly 
create $y$- and $z$-triplons, thus suppressing the formation of ABS as depicted 
in Fig. \ref{fig:4}. Third, symmetry properties of the underlying bilayer 
lattice are respected by $I(\omega)$ as the response is identical for the two 
symmetry related scattering geometries (1) and (4).

\begin{figure}[tb]
\includegraphics[width=1\columnwidth]{
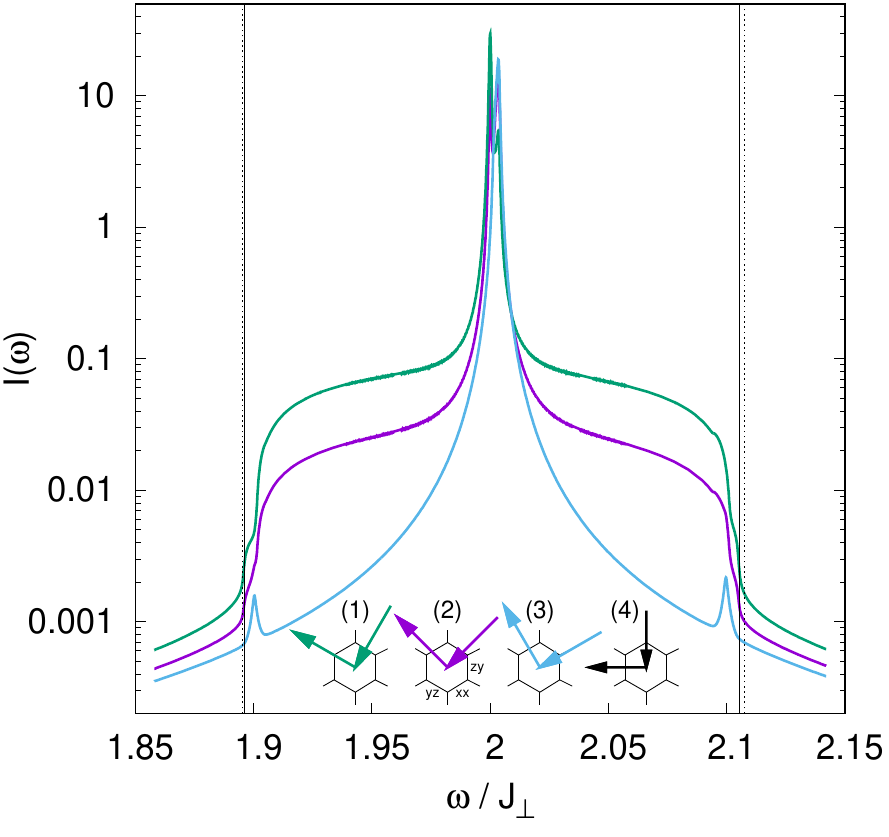}
\caption{Raman intensity $I^0(\omega)$ \textit{without} two-triplon interactions
in $(x,x), (y,y), (z,z)$-sector of AC-stacking at $J = -0.1 J_\perp$ versus
energy transfer $\omega$ for three distinct in-plane, perpendicularly polarized,
scattering geometries, shown by hexagons (1), (2) and (3). Due to the symmetry 
of the model $I^0(\omega)$ is identical for geometries (1) and (4). Letters 
$\alpha\beta$ on hexagon bonds label upper/lower layer $J_{\alpha/\beta}$ 
exchange. Dashed(solid) vertical lines: ABS(continuum boundary) energies.}
\label{fig:8}
\end{figure}

Triplet interactions have a strong impact on the Raman intensity. This can be
seen by contrasting $I(\omega)$ from Fig. \ref{fig:7} against $I^0(\omega)$ from
Fig. \ref{fig:8}. In the latter the irreducible two-triplon entries of $H_{\rm
eff}$ have been turned off. As is obvious, two of the five peaks in $I(\omega)$
are absent from $I^0(\omega)$, and two are strongly suppressed, implying that
all of the former are direct consequences of two-triplon
interactions. Moreover, the variation with polarization angle is markedly
different. In particular no suppression of the central peak occurs, as for
$I(\omega)$.

Regarding the ABS, Raman scattering on the KHBM turns out to behave very
different from conventional $SU(2)$ bilayer spin-systems. In fact, the most
prominent ABS, i.e., those from the $(y,z)$ sector remain hidden from light
scattering. This is because the Raman operator
Eq. (\ref{eq:7}) will not create two-triplon states of this
spin character. Instead, $\mathcal{R}$ can address only states from the $(x,x),
(y,y), (z,z)$-sector, which, as discussed in Sec. \ref{sec:twopart:iso}, are
split-off from the continuum only very weakly. This can be seen both, in the
bound and antibound energy range of Fig. \ref{fig:7}. Here the Raman intensity
displays a fine structure of {\em two} peaks. One of them corresponds to the
upper edge of the continuum, the other one refers to the ABS. This would
certainly pose a challenge to experimental observation of such
states. Remarkably, occurrence of the ABS in the Raman spectrum depends on the
scattering geometry. This can be seen in Fig. \ref{fig:7}, with configuration
(3) not displaying ABS.

\section{Conclusion}
\label{sec:conclusion}

To summarize, using series expansion in terms of the perturbative Continuous
Unitary Transformation, we have investigated the elementary excitations of
the Kitaev-Heisenberg bilayer magnet in its quantum-dimer phase, with a
particular focus on the two-triplon spectrum. In contrast to conventional,
$SU(2)$-invariant bilayer quantum-magnets, we have shown the
Kitaev-Heisenberg bilayer to display a diverse structure of elementary
excitations, depending on two aspects, i.e., the stackings of the bilayer and
the triplon components.

Displaying evidence from two sources, i.e., the spectrum and the wave
functions, we have disclosed several collective (anti)bound two-triplon
states. In principle, we find them to exist for all stackings, however not
for all combinations of triplon components. Moreover, for the two-triplon
(anti)binding energies, we observe a strong variation versus the type of
stacking, with the most prominent (anti)bound states to occur in the
AC-stacking. For some of the (anti)bound states we have provided a simple,
low-order local picture, involving NN-intralayer Kitaev exchange-links, to
explain the (anti)binding of two triplons. This picture is corroborated
by our pCUT calculations, using anisotropic exchange couplings.

For an observable probe of our results, we have calculated the magnetic
Raman response, resulting from a Loudon-Fleury scattering off the Kitaev
planes. We found the Raman intensity to be strongly modified by the
two-triplon interactions. In addition to that, we have identified
Raman-signatures of (anti)bound states close to the continuum. However,
unfortunately, the Raman operator exhibits a selection rule which forbids
scattering from the most prominent of the (anti)bound states. Finally we
have shown the global shape of the Raman intensity to be very sensitive to
the scattering geometry.

\begin{acknowledgments}
This work has been supported in part by the DFG through Project A02 of SFB
1143 (Project-Id 247310070). Work of W.B. has been supported in part by
Nds. QUANOMET (project NP-2), and by the National Science Foundation under
Grant No. NSF PHY-1748958. W.B. also acknowledges kind hospitality of the
PSM, Dresden.
\end{acknowledgments}

\appendix
\section{pCUT-spectra for the KHBM}
\label{sec:appendix:spectrum}

In this appendix we clarify some details specific to the application of the
pCUT to the KHBM. A general description of the method is provided in Ref. 
\cite{Knetter2000a}. The main idea is to block-diagonalize the Hamiltonian into 
an effective Hamiltonian $H_{\mathrm{eff}}$. Each block encodes mixing of
states corresponding to only a particular number $Q$ of excitation quanta,
i.e., $[H_{\text{eff}},H_0] = 0$. Each block can be treated separately.

The interacting groundstate energy can be extracted from the $Q=0$ block.
For the KHBM the unperturbed groundstate is a unique
singlet product state, i.e., the $Q=0$ block has size $1\times1$ and the
groundstate energy can be read off directly from the single matrix element
obtained by pCUT.

The one-triplon dispersion results from diagonalizing the $Q=1$
block. Due to translational invariance of the Hamiltonian, the first step of
this follows from Fourier transformation, which yields a momentum
dependent $6\times 6$ matrix ${\bf E}_{\mathbf{k},jl,\alpha\beta} =
\sum_{\bf r} e^{i\br_j \cdot {\bf k} } \langle \br_j\alpha | H_{\rm eff} |
{\bf 0}_l \beta \rangle - \delta_{ \br_j {\bf 0}_l} \delta_{\alpha\beta}
E_0^{\text{cl}}$ with $\br_{j=1,2} = (\mathbf{r},\mathbf{r}+\boldsymbol{
\delta}^1_x ), \alpha,\beta=x,y,z$ and $E_0^{\text{cl}}$ evaluated on the
same cluster geometry as the corresponding $\langle {\bf 0}_l \alpha |
H_{\rm eff} | {\bf 0}_l \alpha \rangle$ to obtain the irreducible
one-triplon matrix element. The generation of clusters is clarified in App.
\ref{sec:appendix:cluster}.

The preceding $6\times 6$ matrix splits into three $2\times 2$ matrices,
because the effective Hamiltonian $H_{\text{eff}}$ conserves the parity
$P_\alpha$ of the number of each triplon component $\alpha = x,y,z$, i.e.,
\begin{equation}
P_\alpha = e^{i\pi\sum_{\br_j}\hat{n}_{\br_j}^\alpha}\,.
\label{eq:a1}
\end{equation}
Here $\hat{n}_{\br_j}^\alpha$ is the triplon number of triplon $\alpha$ on
site $\br_j$. All $3\cdot2$ triplon-dispersions
$E_{1,2}^\alpha(\mathbf{k})$ can therefore be obtained analytically, and
are described in the main text and in Ref. \cite{Seifert2018}.

To proof the parity conservation, one has to analyze the effective spin
exchange within pCUT. This can be achieved conveniently by using the
singlet-triplon basis as in Eq. \eqref{eq:2} and expressing the
spin-components via singlet and triplon bond operators \cite{Sachdev1990}
\begin{equation}
S_{\br_j,L}^\alpha = \pm t_{\br_j}^{\alpha\,\dagger} 
s_{\br_j}^{\phantom{\dagger}} \pm s_{\br_j}^\dagger 
t_{\br_j}^\alpha - i \epsilon_{\alpha\beta\gamma} 
t_{\br_j}^{\beta\,\dagger} t_{\br_j}^\gamma
\end{equation}
where $\alpha = x,y,z$ is the spin-component, $+(-)$ refers to layer
$L=1(2)$, and $t_{\br_j}^{\alpha\,(\dagger)}$ ($s_{\br_j}^{(\dagger)}$)
creates (annihilates) $\alpha$-triplons (singlets) on the interlayer-dimer
at site $\br_j$.  These operators fulfill bosonic commutator relations and
need to satisfy the constraint
\begin{equation}
s_{\br_j}^{\dagger} 
s_{\br_j}^{\phantom{\dagger}}+t_{\br_j}^{\alpha\,\dagger}
t_{\br_j}^{\alpha\,} = 1
\end{equation}
to suppress nonphysical states. The Kitaev reads
\begin{equation}
S_{\br_{j},L}^\alpha S_{\br'_{j'},L}^\alpha = \sum_{n=-2}^{+2} 
T^{n}_{\br_{j},\br'_{j'},L,\alpha}
\end{equation}
where $\br^{(\prime)}_{j^{(\prime)}}$ are NN sites on layer
$L$, and the ladder operators $T^{n}_{\br_{j},\br'_{j'},L,\alpha}$
non-locally (in)decrement the number of triplons by $n$ at sites $\br^{ (
\prime ) }_{ j^{ ( \prime ) }}$, i.e.,
\begin{align}
T^{0}_{\br_{j},\br'_{j'},L,\alpha} &= 
s_{\br_{j}}^\dagger t_{\br'_{j'}}^{\alpha\,\dagger} 
t_{\br_{j}}^{\alpha} s_{\br'_{j'}}^{\phantom{\dagger}}
+t_{\br_{j}}^{\alpha\,\dagger} s_{\br'_{j'}}^\dagger  
s_{\br_{j}}^{\phantom{\dagger}} t_{\br'_{j'}}^{\alpha}
\label{eq:a5}
\\
&\phantom{=}+(\epsilon_{\alpha\beta\gamma})^2
\left(t_{\br_{j}}^{\beta\,\dagger} t_{\br'_{j'}}^{\gamma\,\dagger} 
t_{\br_{j}}^{\gamma} t_{\br'_{j'}}^{\beta}
-t_{\br_{j}}^{\beta\,\dagger} t_{\br'_{j'}}^{\beta\,\dagger} 
t_{\br_{j}}^{\gamma} t_{\br'_{j'}}^{\gamma}\right)\,,
\nonumber
\\
T^{+1}_{\br_{j},\br'_{j'},L,\alpha} &= \mp i 
\epsilon_{\alpha\beta\gamma} \left( 
t_{\br_{j}}^{\alpha\,\dagger}t_{\br'_{j'}}^{\beta\,\dagger} 
s_{\br_{j}}^{\phantom{\dagger}} t_{\br'_{j'}}^{\gamma} + 
t_{\br_{j}}^{\beta\,\dagger}t_{\br'_{j'}}^{\alpha\,\dagger} 
t_{\br_{j}}^{\gamma} s_{\br'_{j'}}^{\phantom{\dagger}} \right)
\nonumber
\\
&= (T^{-1}_{\br_{j},\br'_{j'},L,\alpha})^\dagger
\label{eq:a6}\,,
\\
T^{+2}_{\br_{j},\br'_{j'},L,\alpha} &= 
t_{\br_{j}}^{\alpha\,\dagger} t_{\br'_{j'}}^{\alpha\,\dagger} 
s_{\br_{j}}^{\phantom{\dagger}} s_{\br'_{j'}}^{\phantom{\dagger}} = 
(T^{-2}_{\br_{j},\br'_{j'},L,\alpha})^\dagger\ .
\label{eq:a7}
\end{align}
The ladder operators satisfy $[T^{n}_{\br_{j},\br'_{j'},L,\alpha},H_0]= n\,
T^{n}_{\br_{j},\br'_{j'},L,\alpha}$ and
$[T^{n}_{\br_{j},\br'_{j'},L,\alpha},P_\beta] = 0$ for $n=0,\pm 2$, as well
as $\lbrace T^{n}_{\br_{j},\br'_{j'},L,\alpha},P_\beta\rbrace = 0$ for
$n=\pm 1$.

Now, turning to Eq. (\ref{eq:3}), each $C_{l,m,n}$, at $O(k=l{+}m{+}n)$ of the
pCUT series for the effective Hamiltonian comprises a real numbered weight
from the flow equations \cite{Knetter2000a} and an operator
\begin{equation}
T(\lbrace n_i\rbrace) = \prod_{i=1}^{k}
T^{n_i}_{\br_{j_i},\br'_{j'_i},L_i,\alpha_i}
\end{equation}
with $n_i \in \lbrace-2,-1,0,+1,+2\rbrace$. $Q$-conservation implies
$\sum_{i=1}^{k} n_i = 0$ and therefore each $T(\lbrace n_i\rbrace)$ may
contain only an even number of $n_i = \pm 1$ ladder operators. In turn
$[T(\lbrace n_i\rbrace),P_\alpha] = 0$ for all $T(\lbrace n_i\rbrace)$.
This proves the parity conservation of the pCUT.

\begin{figure}
$\left(\includegraphics[valign=c,width=0.15\textwidth]{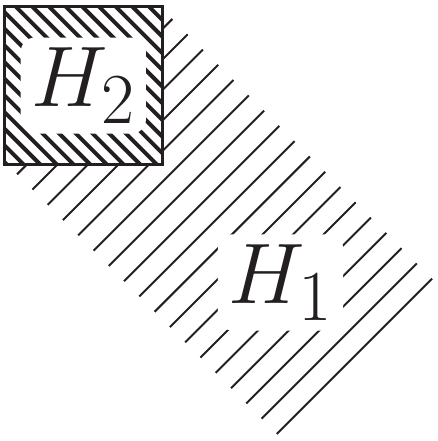}\right)$
\caption{Visualization of the structure of the $Q=2$ block of the
Hamiltonian \eqref{eq:1} for some fixed total momentum $\mathbf{K}$.
When ordered by the relative two-triplon distance $\vert\mathbf{d}\vert$, 
the two-triplon interactions $H_2$, limited to finite distances by the order 
of the pCUT expansion, form a block in the upper left and the single triplon 
propagation $H_1$ forms a semi-infinite tail of the matrix.}
\label{fig:9}
\end{figure}

The two-triplon spectrum is obtained from the $Q=2$ block.  States in
this block are denoted by $\ket{\br_j, \br_j+ \boldsymbol{\tau}_l, \alpha
\beta}$ where $\br_j$ and $\br_j+ \boldsymbol{\tau}_l \neq \br_j$ are two
distinct sites of the lattice and $\alpha,\beta$ mark the two-triplon
components which are present on the dimers. To avoid double counting, we
introduce an ``ordering''. For that, we choose to have the first triplon
located to the lower left of the second triplon and write ``$\br_j < \br_j+
\boldsymbol{\tau}_l$''. The matrix elements of the two-triplon irreducible
Hamiltonian $H_2$, i.e., that by virtue of which {\em both} triplons in the
$Q=2$ sector move, are obtained by subtracting off the two-triplon
reducible contributions from the matrix elements of $H_{\text{eff}}$, which
decomposes into $H_2 + H_1 + H_0$, where $H_{1(0)}$ move at most $1(0)$
triplons \cite{Knetter2003}, i.e.,
\begin{align}
&\bra{\tilde{\br}^{\phantom{\prime}}_{\tilde{j}}, \tilde{\br}'_{\tilde{j}}, 
\alpha \beta}H_2\ket{\br^{\phantom{\prime}}_j, \br'_{j'}, \alpha' \beta'} =
\\ \nonumber
&\bra{\tilde{\br}^{\phantom{\prime}}_{\tilde{j}}, \tilde{\br}'_{\tilde{j}}, 
\alpha \beta}H_{\text{eff}}-H_1-H_0\ket{\br^{\phantom{\prime}}_j, \br'_{j'}, 
\alpha' 
\beta'} =
\\ \nonumber
&\phantom{-}a^{cl}_{\tilde{\br}^{\phantom{\prime}}_{\tilde{j}}, 
\tilde{\br}'_{\tilde{j}},\alpha \beta,
\br^{\phantom{\prime}}_j, \br'_{j'}\alpha' 
\beta'} - E_0^{\text{cl}} \delta_{\br^{\phantom{\prime}}_j,
\tilde{\br}^{\phantom{\prime}}_{\tilde{j}}} 
\delta_{\br'_{j'},\tilde{\br}'_{\tilde{j}}}
\delta_{\alpha\alpha'}\delta_{\beta\beta'}
\\ \nonumber
&-t^{\text{cl}}_{\br^{\phantom{\prime}}_j, 
\tilde{\br}^{\phantom{\prime}}_{\tilde{j}},\alpha}\delta_{\br'_{j'},
\tilde{\br}'_{\tilde{j}}} 
\delta_{\alpha\alpha'}\delta_{\beta\beta'}
-t^{\text{cl}}_{\br'_{j'}, 
\tilde{\br}'_{\tilde{j}},\beta}\delta_{\br^{\phantom{\prime}}_j,
\tilde{\br}^{\phantom{\prime}}_{\tilde{j}}} 
\delta_{\alpha\alpha'}\delta_{\beta\beta'}
\\ \nonumber
&-t^{\text{cl}}_{\br^{\phantom{\prime}}_j, 
\tilde{\br}'_{\tilde{j}},\alpha}\delta_{\br'_{j'},
\tilde{\br}^{\phantom{\prime}}_{\tilde{j}}} 
\delta_{\alpha\beta'}\delta_{\beta\alpha'}
-t^{\text{cl}}_{\br'_{j'}, 
\tilde{\br}^{\phantom{\prime}}_{\tilde{j}},\beta}
\delta_{\br^{\phantom{\prime}}_j,\tilde{\br}'_{\tilde{j}}} 
\delta_{\alpha\beta'}\delta_{\beta\alpha'}\,,
\hphantom{aa}
\end{align}
where $a^{cl}_{\tilde{\br}^{\phantom{\prime}}_{\tilde{j}}, \tilde{ \br
}'_{\tilde{j}},\alpha \beta,\br^{\phantom{\prime}}_j, \br'_{j' } \alpha'
\beta'}$ is the reducible matrix element of $H_{\text{eff}}$ from pCUT, $t^{
\text{cl} }_{ \br^{ \phantom{\prime}}_j, \tilde{ \br }^{
\phantom{\prime}}_{\tilde{j}},\alpha}$ are the irreducible one-triplon 
contributions, $E_0^{\text{cl}}$ is the groundstate energy, and the
superscript $cl$ implies that all of the latter have to be evaluated on the
same cluster.

We obtain the two-triplon energies by evaluating the spectrum of
$H_{\text{eff}} - H_0 = H_1 + H_2$. This can be achieved in two
steps. First, we switch to center of mass and relative coordinates,
$\mathbf{R} = \br_j + \frac{\boldsymbol{\tau}_l}{2}$ and $\mathbf{d} =
\boldsymbol{\tau}_l$, respectively, and introduce the Fourier transform with
respect to $\mathbf{R}$, i.e.,
\begin{equation}
\ket{\mathbf{K},\mathbf{d},\alpha\beta} = \sum_{\mathbf{R},\mathbf{d}}
e^{-i\mathbf{K}\cdot\mathbf{R}} \ket{\mathbf{R}{-}\mathbf{d}/2, 
\mathbf{R}{+}\mathbf{d}/2, \alpha \beta}\,.
\end{equation}
The states $\ket{\mathbf{K},\mathbf{d},\alpha\beta}$ pre-diagonalizes the
$Q=2$ block in terms of the total two-triplon momentum $\mathbf{K}$.

For the remaining diagonalization of $H_1 + H_2$ with respect to the
relative coordinate $\mathbf{d}$, it is instructive to visualize the
structure of the $Q=2$ block at fixed $\mathbf{K}$, shown in
Fig. \ref{fig:9}. It consists of two types of entries obtained from
pCUT. First, a block of irreducible two-triplon interactions $H_2$. The rank
of this, i.e., the {\em range} of the interaction, is not only related to the
order of the pCUT, but the absolute size of its matrix elements versus ${\bf
d}$ also encodes the physical range and spacial structure of the effective
two-triplon interactions. Second, the block displays a semi-infinite banded
tail, encoding the scattering states with no mutual interactions, created by
$H_1$. If $H_2$ is set to zero, the remaining $H_1$ can be diagonalized
{\em analytically} by a second Fourier transform with respect to ${\bf d}$,
yielding a bare two-triplon spectrum $E_i^\alpha(\mathbf{k}) +
E_j^\beta(\mathbf{k}')$ such that $\mathbf{K} = \mathbf{k} + \mathbf{k}'$
and $i,j = 1,2$.

If $H_2$ is finite, it is not possible to analytically diagonalize the
matrix - even though all of its entries are analytic expressions, derived
from the pCUT. Instead the two-triplon scattering problem at hand is
diagonalized numerically, choosing a lattice of linear dimension $\hat{d}$,
large enough to contain all of $H_2$ at a given order of the pCUT and
scattering states up to a distance large enough, to reduce finite size
effects sufficiently. As shown in Fig. \ref{fig:5}, we find that at $O(8)$,
a $\hat{d}\simeq 8$ is sufficient, both, for a faithful representation of
the two-triplon continuum and the ABS.

As a final simplification, parity conservation for triplon numbers,
Eq. (\ref{eq:a1}), also applies in the $Q=2$ block. It splits the $H_1 +
H_2$ matrix into four subblocks which can be diagonalized
independently. These are three separate subblocks with triplon components
$\lbrace(x,y)\rbrace$, $\lbrace(y,z)\rbrace$ and $\lbrace(z, x)\rbrace$
only, and a fourth block mixing two triplons with components
($\lbrace(x,x)$, $(y,y)$, $(z, z)\rbrace$).  In the latter all triplon
numbers are of even parity while in the former two triplon numbers have an
odd parity.

\section{Cluster generation}
\label{sec:appendix:cluster}

In this Appendix we describe the generation of clusters on which pCUT matrix
elements are evaluated. By virtue of the {\em linked cluster theorem}, pCUT
hopping matrix elements, in order to be size-consistent, have to be calculated
at least on the largest cluster, that is linked through the effective
Hamiltonian at a given order $N$ of the series expansion \cite{Kadanoff1981,
Marland1981}. Since calculation times grow exponentially with cluster size,
it is important to generate the clusters such as to include only those
lattice sites and bonds which are linked at most.

The cluster generation takes slightly different routes for one-triplon,
two-triplon, and Raman operator matrix elements. Each of them is described
briefly in the following.

\subsection{Cluster for one-triplon matrix elements}

A one-triplon matrix element $\langle \br_j\alpha | H_{\rm eff} | {\bf 0}_l
\beta \rangle$ represents the transition amplitude between an initial state
$\ket{{\bf 0}_l \beta}$ with exactly one triplon at some initial site ${\bf
0}_l$ in a singlet-background on all other sites and a final state
$\ket{\br_j\alpha}$ with the triplon at some final site $\br_j$. In the
intermediate states of the effective Hamiltonian all triplon flavors
mix. Therefore, a classification of the cluster with respect to the latter
is not performed.

To generate the cluster for the matrix element $\langle \br_j | H_{\rm eff}
| {\bf 0}_l \rangle$ one has to include all possible paths $P^{k',i}_{{\bf
0}_l, \br_j}$ the triplon can take to transit from site ${\bf 0}_l$ to site
$\br_j$, where $0\leq k'\leq k$, $k$ is the expansion order and $i$ is an
index to enumerate the paths.  $P^{k,i}_{{\bf 0}_l, \br_j}$ is a set of site
pairs $(\br_i,\br'_{i'})$ where each pair encodes a single-order move
of the triplon across the lattice. Since the KHBM comprises only
NN-exchange, the triplon can, at most, move to an adjacent NN site per order
of the pCUT, resulting in $k$ NN-hoppings for a $k$-th order expansion,
limiting the cardinality of each $P^{k,i}_{{\bf 0}_l, \br_j}$ strictly to
$k$.

\begin{figure}[tb]
\includegraphics[width=0.25\textwidth]{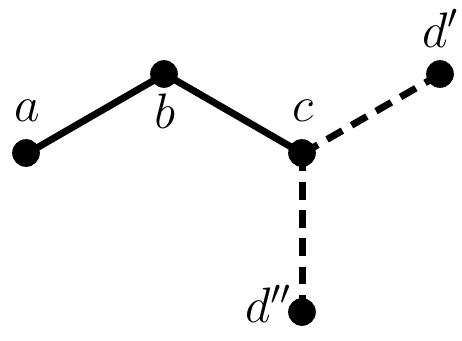}
\caption{One step in the construction of 1-triplon paths. The second order 
path $P^{2}_{a,c} = \{(a,b),(b,c)\}$ can be extended to three distinct third
order paths $P^{3,1}_{a,d'} = \{(a,b),(b,c),(c,d')\}$, $P^{3,2}_{a,d''} = 
\{(a,b),(b,c),(c,d'')\}$ and $P^{3,3}_{a,b} = \{(a,b),(b,c),(c,b)\}$.}
\label{fig:10}
\end{figure}

All paths $P^{k,i}_{{\bf 0}_l, \br_j}$ up to $k$-th order are obtained
iteratively in combination with simultaneous order-by-order calculation of
{\itshape all} matrix elements. I.e., by hopping a triplon to NN sites,
starting at the final site of a $k$-th order path, $k{+}1$-th order paths
are generated
\begin{equation}
P^{k+1,i'}_{{\bf 0}_l, \br'_{j'}=\br_j\pm\delta_\alpha^1} = P^{k,i}_{{\bf 0}_l, 
\br_j} \cup \{(\br_j,\br_j\pm\delta_\alpha^1)\}\,.
\end{equation}
The index $i'$ enumerates the new paths generated. The sign
$\pm$ is chosen such that $\br_j\pm\delta_\alpha^1$ is a valid lattice site.
See Fig. \ref{fig:10} for an example.

From the preceding, the minimal linked cluster $C^k_{{\bf 0}_l, \br_j}$,
required to calculate a matrix element $\langle \br_j | H_{\rm eff} | {\bf
0}_l \rangle$ to order $k$ is obtained by combining the sites and bonds of
all paths $P^{k',i}_{{\bf 0}_l, \br_j}$ connecting the two sites, i.e.,
\begin{equation}
	C^k_{{\bf 0}_l, \br_j} = \bigcup_{k'\leq k,i} P^{k',i}_{{\bf 0}_l, \br_j}\ .
\end{equation}
For the KHBM we use rotation-symmetries of the lattice to reduce the
number of clusters to be generated.

If a cluster $C^k_{{\bf 0}_l, \br'_j}$ is a subset of another cluster (with 
identical initial sites), i.e.,
\begin{equation}
	C^k_{{\bf 0}_l, \br'_j} \subset C^k_{{\bf 0}_l, \br_j}\ ,
\end{equation}
we use a single application of pCUT on the larger cluster $C^k_{{\bf 0}_l, 
\br_j}$ to calculate both matrix elements simultaneously, saving significant 
computation time.

\subsection{Cluster for two-triplon matrix elements}

Extending the preceeding, the two-triplon matrix elements 
$\bra{ \tilde{ \br }^{ \phantom{ \prime }}_{ \tilde{j}}, \tilde{\br}'_{\tilde{j}}, 
\alpha \beta}$ $H_{\text{eff}}$ $\ket{\br^{\phantom{\prime}}_j, \br'_{j'}, \alpha' 
\beta'}$ represent transition amplitudes
between an initial state $\ket{\br^{\phantom{\prime}}_j, \br'_{j'}, \alpha' 
\beta'}$ with exactly two triplons at initial
sites $\br^{\phantom{\prime}}_j$ and $\br'_{j'}$ in a singlet-background on all 
other sites and a final state $\ket{\tilde{\br}^{\phantom{\prime}}_{\tilde{j}}, 
\tilde{\br}'_{\tilde{j}},\alpha\beta}$ with the triplons at final sites 
$\tilde{\br}^{\phantom{\prime}}_{\tilde{j}}$ and $\tilde{\br}'_{\tilde{j}}$.
Again, the triplon flavor is not considered.

\begin{figure}[tb]
\includegraphics[width=0.25\textwidth]{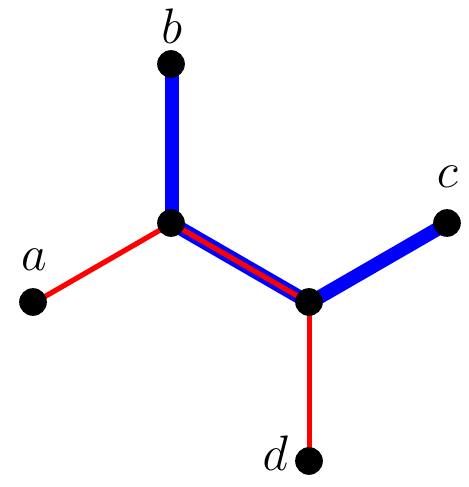}
\caption{Construction of a two-triplon path: the combination
of two third order one-triplon paths $P^{3,i}_{\mathbf{a},\mathbf{d}}$ 
(red) and $P^{3,i'}_{\mathbf{b},\mathbf{c}}$ (blue) 
result in a fifth order two-triplon-path 
$P^{5,i''}_{\mathbf{a}\mathbf{b},\mathbf{c}\mathbf{d}}$ with one shared bond 
and two shared sites. This path will be kept for the cluster generation.}
\label{fig:11}
\end{figure}

The generation of clusters for two-triplon matrix elements
$\bra{\tilde{\br}^{\phantom{\prime}}_{\tilde{j}}, \tilde{\br}'_{\tilde{j}}}
H_{\text{eff}} \ket{\br^{\phantom{\prime}}_j, \br'_{j'}}$ makes use of the
one-triplon-paths $P^{k,i}_{{\bf 0}_l, \br_j}$ generated in the previous
section.  Any linked cluster
$P^{k'',i''}_{\br^{\phantom{\prime}}_j\br'_{j'},
\tilde{\br}^{\phantom{\prime}}_{\tilde{j}} \tilde{\br}'_{\tilde{j}}}$,
allowing for two-triplon irreducible interactions comprises two one-triplon
paths $P^{k,i}_{ \br^{\phantom{ \prime}}_j, \tilde{\br}^{ \phantom{\prime
}}_{ \tilde{j}}}$ and $P^{k',i'}_{\br'_{j'}, \tilde{\br}'_{\tilde{j}}}$, or
those with the final sites exchanged, such that the paths have (at least) a
common site. See Fig. \ref{fig:11} for an example. I.e.,
\begin{equation}
	P^{k'',i''}_{\br^{\phantom{\prime}}_j\br'_{j'}, 
	\tilde{\br}^{\phantom{\prime}}_{\tilde{j}} \tilde{\br}'_{\tilde{j}}} = 
	P^{k,i}_{\br^{\phantom{\prime}}_j,\tilde{\br}^{\phantom{\prime}}_{\tilde{j}}
	} \cup P^{k',i'}_{\br'_{j'}, \tilde{\br}'_{\tilde{j}}}
\end{equation}
where $k''$ is the order (read number of bonds) of the two-triplon path and 
$i''$ is an appropriate index.
By construction the order of the new path has to fulfill $\max(k,k')\leq k'' 
\leq k+k'$.

Analogous to the one-triplon case the minimal $k$-th order two-triplon 
linked cluster $C^k_{\br^{\phantom{\prime}}_j\br'_{j'}, 
\br^{\phantom{\prime}}_{\tilde{j}} \tilde{\br}'_{\tilde{j}}}$ is obtained by 
combining all two-triplon paths with irreducible two-triplon interactions, 
i.e.,
\begin{equation}
	C^k_{\br^{\phantom{\prime}}_j\br'_{j'}, 
	\tilde{\br}^{\phantom{\prime}}_{\tilde{j}} \tilde{\br}'_{\tilde{j}}} = 
	\bigcup_{k''\leq k,i''} P^{k'',i''}_{\br^{\phantom{\prime}}_j\br'_{j'}, 
	\tilde{\br}^{\phantom{\prime}}_{\tilde{j}} \tilde{\br}'_{\tilde{j}}}\ .
\end{equation}

\subsection{Cluster for two-triplon states generated by $\mathcal{R}_{\rm eff}$}
\label{ss:Reffcl}

To determine the Raman response, a third type of cluster is needed to
evaluate $\mathcal{R}_{\rm eff}\ket{}$ from Eq. \eqref{eq:8}. 
As we confine ourselves to the $Q=2$ contribution from $\mathcal{R}_{\rm 
eff}$ it effectively creates a pair of triplons from the singlet-product state 
and distributes it across the lattice, generating an effective excitation cloud.

Because the Loudon-Fleury vertex from Eq. \eqref{eq:7} is translational 
invariant and only contains local spin interactions, it is sufficient to treat 
the Raman operator as acting on a single bond positioned at $({\bf 0}, 
\boldsymbol{\delta}_\alpha)$ only. The clusters are constructed from pseudo 
two-triplon-paths by combining two one-triplon paths 
$P^{k,i}_{\br^{\phantom{\prime}}_j,\tilde{\br}^{\phantom{\prime}}_{\tilde{j}}}$
and $P^{k',i'}_{\br'_{j'}, \tilde{\br}'_{\tilde{j}}}$, requiring that one
path must include the first site of the Raman bond and the second one the
second site, i.e.,
\begin{equation}
{\bf 0}\in P^{k,i}_{\br^{\phantom{\prime}}_j,
\tilde{\br}^{\phantom{\prime}}_{\tilde{j}}}
\quad\text{and}\quad
\boldsymbol{\delta}_\alpha \in P^{k',i'}_{\br'_{j'}, 
\tilde{\br}'_{\tilde{j}}}\ .
\end{equation}
Furthermore, in contrast to the two-triplon case, both paths do {\it not}
need to share a site, as this occurs via the Raman bond.  Finally, the two
starting sites $\br^{\phantom{\prime}}_j$ and $\br'_{j'}$ need to form a
bond included in at least one of both paths or being the Raman bond 
$({\bf 0}, \boldsymbol{\delta}_\alpha)$.  The former ensures that the
Raman operator can interact with both triplons, the latter mimics the
creation of both triplons from the ground state.

The Raman clusters are ultimately formed by combining all valid two-path 
configurations that lead to the same final positions
$\tilde{\br}^{\phantom{\prime}}_{\tilde{j}}$ and $\tilde{\br}'_{\tilde{j}}$
regardless of the initial ones.

If, as in the present case, the observable is a sum of operators on different 
types of bonds, each of them requires generating a separate set of clusters. 
Their contributions to $\mathcal{R}_{\rm eff}\ket{}$ are additive.

\section{Evaluation of the Raman intensity}
\label{sec:appendix:raman}

This appendix explains details for the evaluation of the Raman intensity,
i.e., Eq. \eqref{eq:8}.
The Loudon-Fleury vertex, see Eq. \eqref{eq:7}, can be rewritten as
\begin{equation}
	\mathcal{R} = \sum_{\mathbf{r}} \mathcal{R}(\mathbf{r})
\end{equation}
with $\mathcal{R}(\mathbf{r})$ acting only on the spins at site $\mathbf{r}$ 
and its NN-sites. Applying the pCUT yields the effective Raman operator
\begin{equation}
	\mathcal{R}_{\text{eff}} = \sum_{\mathbf{r}} 
	\mathcal{R}_{\text{eff}}(\mathbf{r})
\end{equation}
with $\mathcal{R}_{\text{eff}}(\mathbf{r})$ now acting on a finite portion of 
the lattice around the site $\mathbf{r}$, depending on the order of the series 
expansion. From this we obtain the effective Fourier transformed Raman operator
\begin{equation}
	\mathcal{R}_{\text{eff}}(\mathbf{K}) = \frac{1}{\sqrt{N}}\sum_{\mathbf{r}} 
	\mathrm{e}^{i\mathbf{K}\mathbf{r}} \mathcal{R}_{\text{eff}}(\mathbf{r})
\end{equation}
where $\mathbf{K}$ is the momentum transferred and $N$ is the number of lattice 
sites.

As mentioned in Sec. \ref{sec:raman}, $\mathcal{R}_{\rm eff}$ is not diagonal
in $Q$ and we focus on the lowest number of triplons excited only, i.e.,
$Q=2$.  Thus we confine $\mathcal{R}_{\text{eff}}(\mathbf{r})$ to those parts
$\mathcal{R}_{\text{eff},2}(\mathbf{r})$ that excite exactly two triplons
from the ground state $\ket{}$. Because $\mathcal{R}_{\text{eff}}(\mathbf{r})$ 
only acts on a finite region of the lattice the resulting state 
$\mathcal{R}_{\text{eff},2}(\mathbf{r})\ket{}$ comprises a finite sum of 
two-triplon states $\ket{\br_j, \br_j+\boldsymbol{\tau}_l, \alpha \beta}$ only. 
Thus $\mathcal{R}_{\text{eff},2}(\mathbf{K})\ket{}$ will also contain only a 
finite number of two-triplon states $\ket{\mathbf{K}, \boldsymbol{ 
\tau}_l,\alpha\beta}$ of fixed momentum.

This allows to treat the action of 
$\left[\omega+i\,0^+-H_{\text{eff}}+E_0\right]^{-1}$ in Eq. \eqref{eq:8} by 
tridiagonalization \cite{Knetter2003, Lanczos1950, Viswanath1994}, resulting in a continued fraction expression for the Raman 
intensity, i.e.,
\begin{equation}
	-\pi I(\omega) = {\rm Im} 
	\frac{\bra{}\mathcal{R}_{\text{eff},2}^\dagger(\mathbf{K})\mathcal{R}_{
	\text{eff},2}(\mathbf{K})\ket{}}{\omega-a_0-\frac{b_1^2}{\omega-a_1-
	\frac{b_2^2}{\omega-\dots}}}\ .
\end{equation}
The coefficients $a_n$, $b_n^2$ are obtained by iteration
\begin{align}
\ket{f_{n+1}} &= (H_{\text{eff}}-E_0)\ket{f_n} -
 a_n\ket{f_n}-b^2_n\ket{f_{n-1}} \nonumber\\
a_n&= \frac{\bra{f_n}(H_{\text{eff}}-E_0)\ket{f_n}}{\langle
 f_n\vert f_n\rangle} \nonumber\\
b^2_{n+1}&= \frac{\langle f_{n+1}\vert f_{n+1}\rangle}{\langle f_n
\vert f_n\rangle}\,, 
\end{align}
starting with $\ket{f_0} = \mathcal{R}_{\text{eff},2}(\mathbf{K})\ket{}$.

The action of the effective Hamiltonian can be replaced by
$H_{\text{eff}}-E_0 = H_1 + H_2$, allowing to reuse matrix elements for
propagating the states $\ket{f_n}$, which have been calculated previously for
the one- and two-triplon spectra (see appendix \ref{sec:appendix:spectrum}).
Because the iteration only requires to save the last three states, the
calculation is very memory efficient and limited only by the computation
time allotted. We evaluate coefficients up to $n=500$ and truncate the
continued fraction at that point by setting $b_{500} = 0$.

\end{document}